\documentclass[12pt]{article}

\usepackage{scicite}
\usepackage{times}

\newcommand{\be}{\begin{equation}}
\newcommand{\ee}{\end{equation}}
\usepackage{amsmath}
\usepackage{amssymb}
\usepackage{physics}
\usepackage{graphicx}
\usepackage{float}
\usepackage[labelfont=bf]{caption}
\usepackage[bookmarks=false,breaklinks=false,pdfborder={0 0 0},pdfborderstyle={},backref=false,colorlinks=false]{hyperref}
\usepackage{appendix}

\newenvironment{sciabstract}{%
\begin{quote} \bf}
{\end{quote}}

\title{Chip-Scale Point-Source Sagnac Interferometer by Phase-Space Squeezing}

\author
{ Yiftach Halevy$^{\dag}$, Yali Cina$^{\dag}$, Omer Feldman, David Groswasser,\\ Yonathan Japha$^{\ast}$ and Ron Folman\\
\\
\normalsize{Department of Physics, Ben-Gurion University of the Negev,}\\
\normalsize{Beer Sheva blvd 1, 84105 Beer Sheva, Israel}\\
\\
\normalsize{$^\ast$ Corresponding author. E-mail:  japhay@bgu.ac.il.}\\
\normalsize{$^{\dag}$ These authors contributed equally to this work}
}

\date{}

\begin{document}

\baselineskip15pt

\maketitle

\begin{sciabstract}

Matter-wave interferometry plays a significant role in scientific research and technological applications. While position-momentum phase-space squeezing has been demonstrated to increase the coherence of atom sources by reducing momentum spread, we theoretically investigate the potential advantages of the opposite squeezing. As a case study, we analytically and numerically examine its effect on point source atom interferometry (PSI) for rotation sensing. Our analysis reveals that this squeezed PSI (SPSI) approach can significantly improve sensitivity and dynamic range while enabling shorter cycle times and higher repetition rates. Through simulations, we identify parameter spaces where sensitivity and dynamic range are enhanced by orders of magnitude. Under a specific definition of compactness, our calculations show that SPSI outperforms standard PSI by over four orders of magnitude. These theoretical findings suggest that SPSI could either enhance performance in standard-sized devices or maintain performance in miniaturized chip-scale devices, potentially paving the way for new practical applications.
    
\end{sciabstract}

\newpage

\section*{Introduction}

Since the pioneering experiments of atom interferometry\cite{kasevich_atomic_1991,keith_interferometer_1991,carnal_youngs_1991,riehle_optical_1991,cronin_optics_2009}, researchers have explored the measurement of rotation using the Sagnac effect in closed-area atom interferometers\cite{anderson_sagnac_1994}. These investigations have revealed exceptionally high sensitivity, comparable to state-of-the-art optical Sagnac interferometers\cite{garrido_alzar_compact_2019}. Since the early 2000s, these advancements have been driven by potential applications in inertial guidance \cite{savoie_interleaved_2018,geiger_continuous_2016},  geophysics \cite{gillot_stability_2014} and space-based research \cite{bertoldi_aedge_2021,bassi_way_2022,elliott_quantum_2023,abend_technology_2023}. The majority of matter-wave interferometers studied since then are laser pulse atom interferometers (LPAIs) that rely on two-photon Raman transitions for splitting, diverting, and recombining the atomic wave packets \cite{gustavson_rotation_2000,durfee_long-term_2006,canuel_six-axis_2006,gauguet_characterization_2009,stevenson_sagnac_2015,gautier_accurate_2022,janvier_compact_2022}.

The sensitivity of LPAIs to linear acceleration relies on a space-time area between two interferometer arms, while sensitivity to rotations requires that the two arms enclose a real space area. To effectively implement LPAI gyroscopes based on the $\pi/2-\pi-\pi/2$ pulse sequence, it is important to differentiate phase shifts induced by rotations from those caused by accelerations of the carrying platform. This can be achieved by combining the signals from two interferometers with counter-propagating atoms \cite{gauguet_off-resonant_2008,stockton_absolute_2011,berg_composite-light-pulse_2015,yao_calibration_2018}. The point-source interferometer (PSI) is a specific kind of light-pulse atom interferometer \cite{dickerson_multiaxis_2013,chen_single-source_2019,avinadav_rotation_2020} that allows sensing both acceleration and rotation by probing the phase and spatial frequency of a fringe pattern formed by a given internal atomic state in an expanding cloud of cold atoms. The method exploits the correlation between position and velocity generated by the expansion of the cloud throughout the interferometer sequence.

The compact nature of the PSI setup makes it suitable for applications where space is limited, such as in portable devices or space-based research and it has the potential to achieve the long sought-after goal of miniaturizing rotation sensing \cite{savoie_interleaved_2018,yao_calibration_2018,krzyzanowska_matter-wave_2023,jia_dual_2024}. However, the sensitivity and dynamic range of PSI is limited because on the one hand the initial cloud should be as cold as possible to allow it to be confined to a small volume but on the other hand it should have a fast expansion rate in order to quickly achieve a good position-velocity correlation and develop a highly resolvable spatial fringe pattern.

Here, we propose a squeezed point-source interferometer (SPSI) for increasing the dynamic range of the interferometer and its sensitivity or, alternatively, achieving the standard sensitivity with a smaller device, or even enable a chip-scale sensor. Our method is based on adding a stage of pre-acceleration where the initial cloud undergoes position-momentum phase-space squeezing. Phase-space squeezing has been discussed extensively regarding delta-kick cooling \cite{ammann_delta_1997,kovachy_matter_2015,luan_realization_2018,dupays_delta-kick_2021,pandey_atomtronic_2021}, in which the spread in momentum is decreased at the expense of increasing the spread in position. In contrast, here, we propose to increase the spread in momentum while the projected spread in position is reduced, in the sense that if the atomic dynamics is reversed, the atoms concentrate in a volume smaller than their actual initial volume.

Our proposed SPSI sequence consists of several key stages. Initially, a thermal cloud of cold atoms is prepared using standard cooling techniques. This is followed by a novel pre-acceleration phase where the atoms experience a repulsive potential, inducing position-momentum phase-space squeezing. Subsequently, the atoms undergo a conventional three-pulse interferometry sequence $(\pi/2 - \pi - \pi/2)$. Finally, the atomic state is detected, revealing a spatial fringe pattern that encodes rotation information. While our current study focuses on theoretical analysis and numerical simulations of this sequence, it provides a foundation for future experimental implementations.

Through our analytical and numerical studies, we demonstrate that integrating an inhomogeneous repulsive force to accelerate atom motion, prior to the interferometer sequence, has the capability to substantially enhance the operational efficiency and sensitivity of a PSI device. Our calculations show that the figures of merit are enhanced by orders of magnitude, and under a definition of compactness, the enhancement in performance is about four orders of magnitude.

\begin{figure}[ht]
    \centering
    \includegraphics[trim = 105mm 70mm 90mm 65mm, clip, width=1\textwidth]{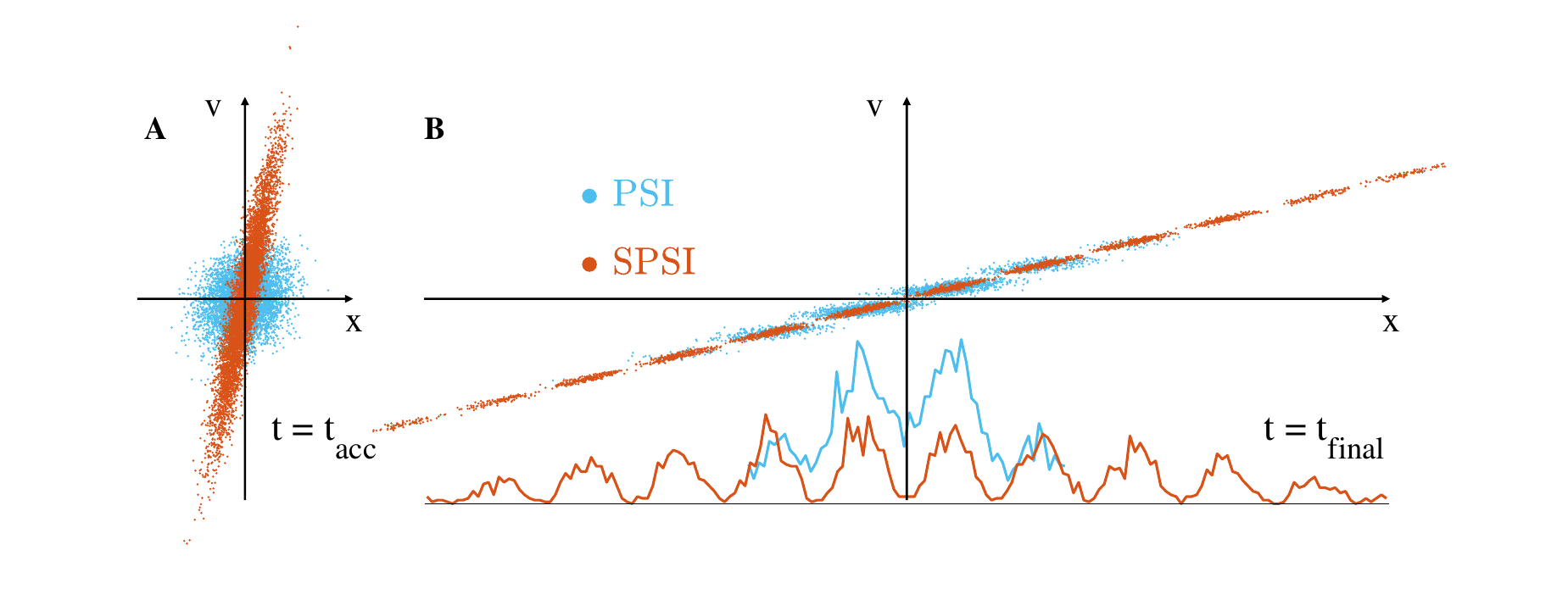}
    \caption{{\bf Principle of the squeezing effect in the interferometer}. The atomic phase-space distribution is shown for the standard PSI interferometer (blue) and the squeezed-PSI (SPSI) interferometer (orange), at two different times: (A) at the start of the interferometer sequence, $t_{\rm{acc}}$, occurring just after the acceleration stage due to the repulsive-potential pulse, and (B) at the imaging time, $t_\text{final}$, following the complete interferometer sequence. The phase-space distribution at $t_{\rm{final}}$ becomes tilted due to expansion, and the velocity-dependent interferometer phase introduces an oscillatory phase $\Delta\phi(v_x)=2k_{\rm eff}\Omega_y T_R^2 v_x\equiv k_v v_x$ that modulates the phase-space distribution as $\rho(x,v_x)\to \frac12 \rho(x,v_x)[1+\cos(k_v v_x)]$ upon detecting a single state. As shown at the bottom of (B), the latter oscillation produces a spatial density oscillation, which, upon being projected onto the position axis, forms the observed signal (output) of the interferometer. In the SPSI, when the initial phase-space distribution is squeezed with increased velocity, the subsequent expansion results in a phase-space distribution having a high aspect ratio. Upon projection onto the position axis, the SPSI signal shows improvement with more oscillations and an improved contrast. This figure was generated using a simplified version of the numerical simulations described in the Materials and Methods section, focusing on illustrating the key principles of phase-space evolution in both PSI and SPSI configurations.}
    \label{fig:Phase-Space Squeezing}
\end{figure}

An example of the effect of phase-space squeezing on the interferometer is illustrated in Fig.\,\ref{fig:Phase-Space Squeezing}. Our simulations indicate that the squeezing leads to both an increase in the number of oscillations within the cloud size and an increase in the contrast.

\section*{Results}

\subsection*{Characteristics and Limits of PSI for Rotation Sensing}

An established method for constructing a Laser-Pulse Atom Interferometry (LPAI) system involves a splitting and recombining laser giving rise to stimulated transitions (e.g., Raman, Bragg, or single photons for clock transitions) to implement a $\pi/2-\pi-\pi/2$ pulse sequence. In an interferometer using freely falling atoms, the interferometer phase shift due to acceleration and rotation of the platform is given by \cite{peters_high-precision_2001,dickerson_multiaxis_2013}

\be \label{eq:LPAI pahse}
\Delta\Phi = \mathbf{K}_{\text{eff}}\cdot\mathbf{a}{T_{R}}^{2}+2\mathbf{K}_{\text{eff}}\cdot\left(\mathbf{\Omega}\times\mathbf{v}\right){T_{R}}^{2},
\ee
where $\mathbf{K}_{\text{eff}}$ is the wave vector of the momentum given to the atom during the laser pulse, $\mathbf{a}$ is the acceleration of the platform relative to the atoms (including gravity), which is assumed to be constant during the measurement, $T_{R}$ is the time interval between the interferometer pulses, $\mathbf{\Omega}$ represents the rotation rate of the platform, and $\mathbf{v}$ denotes the initial velocity of the atom. The second term in this equation is identified as the rotation phase. 

The rotation phase in Eq.\,\ref{eq:LPAI pahse}, can be expressed as the Sagnac phase\cite{sagnac_1914} $\Delta\Phi_{\Omega}=(2m / \hbar)\mathbf{\Omega}\cdot\mathbf{A}$. In this context, $m$ is the atomic mass, while $\mathbf{A}$ is the area enclosed by the interferometer arms. Notably, these two expressions for the phase shift are equivalent, as demonstrated through the relation $\mathbf{A}=\left(\mathbf{v}_{r}T_{R}\right)\times\left(\mathbf{v}T_{R}\right)$, where $\mathbf{v}_{r}=\hbar\mathbf{K}_{\text{eff}}/m$ is the atom's recoil velocity. In the conventional scheme, the momentum transfer in each Raman pulse is equal to the difference of the wave vectors of the two counter-propagating Raman beams, such that $\mathbf{K}_{\text{eff}}\approx 2\mathbf{K}$, where $\mathbf{K}$ is the wave vector of a single Raman laser. Advanced techniques have demonstrated considerably higher values of $\mathbf{K}_{\text{eff}}$ \cite{mcguirk_large_2000,chiow_102_2011,kovachy_quantum_2015,plotkin-swing_three-path_2018,rudolph_large_2020,parker_measurement_2018,gebbe_twin-lattice_2021,li_high_2021,dubetsky_sequential_2023,siems_large-momentum-transfer_2023,goerz_robust_2023,li_spin-squeezing-enhanced_2023,louie_robust_2023}.

In the following we use the point-source interferometer (PSI) as a case study to examine in detail how position-momentum phase-space squeezing enhances the performance of the interferometer. The position and velocity distribution of the atoms prior to and during the interferometer sequence and its effect on performance is the topic of this work. The PSI exploits the correlation between position and velocity generated by the expansion of the cloud throughout the interferometer sequence. According to Eq.\,\ref{eq:LPAI pahse}, the probability that an atom with initial velocity $\mathbf{v}$ is at the end of the interferometric sequence in a given internal state $|s\rangle$ is 

\be \label{eq: state probability}
P_s({\bf v})=1/2\cdot[1+\cos(\mathbf{k}_v\cdot\mathbf{v}+\phi_s)],
\ee
where $\mathbf{k}_v=2(\mathbf{K}_{\rm eff}\times\mathbf{\Omega})T_R^2$ and $\phi_s$ is determined by gravity and acceleration of the platform. In an ideal case, the atomic cloud starts to expand freely from a point source, and the position of each atom is linearly proportional to its velocity $\mathbf{x}=\mathbf{v}T_{\rm{ex}}$, where $T_{\rm{ex}}$ is the time of expansion. In this limit, the Sagnac phase becomes \cite{chen_single-source_2019}

\be \label{eq: PSI rotation phase}
\Delta\Phi_{\Omega,\text{PS}}(\mathbf{x})=\frac{2{T_{R}}^{2}}{T_{\text{ex}}}\left(\mathbf{K}_{\text{eff}}\times\mathbf{\Omega}\right)\cdot\mathbf{x}\equiv\mathbf{k}_{x}\cdot\mathbf{x}.
\ee

In this case, the atoms form a spatial density constituting a fringe pattern that oscillates with a wave-vector $\mathbf{k}_x$, from which one extracts the angular velocity component, transverse to the plane defined by the vectors $\mathbf{K}_{\text{eff}}$ and $\mathbf{k}_x$. It is important to highlight that any deviation from the established position-velocity correlation results in a reduction of contrast within the fringe pattern, consequently diminishing the precision of the rotation measurement. Thus, to attain an approximation of a point source, it is necessary to cool and confine the atoms, resulting in reduced thermal velocities. However, this, in turn, leads to a smaller interferometer area, thus yielding a lower resolution. Therefore, to enhance the resolution of the rotation phase measurement, it is advantageous to increase these velocities while preserving the position-velocity correlation. 

Let us now examine the more general case of an initial atomic cloud with a finite size. We consider a thermal cloud with an initial isotropic Gaussian spatial distribution having a width $\sigma_{x0}$ and a Gaussian velocity distribution having a width $\sigma_{v0}$. In this case, one can show (see supplementary information for an alternative derivation for quantum wave packets) that the output density has the form\,\cite{shahriar_analytical_2017,li_high_2021}

\be \label{eq: Gaussian cloud output density}
\rho_f({\bf x})\propto e^{-x^2/2\sigma_f^2}[1+C\cos(\alpha \mathbf{k}_x\cdot{\bf x}+\phi_s)],
\ee
where $\sigma_f=\sqrt{\sigma_{x0}^2+T_{\rm{ex}}^2\sigma_{v0}^2}$ is the final cloud width, $\alpha=1-\sigma_{x0}^2/\sigma_f^2$, and the contrast $C$ (Eq.\,\ref{eq: Gaussian cloud contrast}) drops considerably when the angular velocity is large such that the fringe period $2\pi/k_x$ becomes as small as the initial cloud size (for $\alpha\sim 1$)

\be \label{eq: Gaussian cloud contrast}
C=\exp\left[-\frac12\alpha k_x^2\sigma_{x0}^2\right]\,.
\ee

The angular velocity $\Omega$ is deduced from the measured atomic density by fitting it to the form of Eq.\,\ref{eq: Gaussian cloud output density} and then extracting $\Omega$ from the periodicity $\alpha k_x$. This periodicity includes a deviation

\be \label{eq: Gaussian cloud deviation}
\Delta\Omega=-\frac{\sigma_{x0}^2}{\sigma_f^2}\Omega\,, 
\ee
which is caused by the non-ideal point-source approximation and may be eliminated to some extent if the ratio between the initial and final cloud sizes is known. The appearance of the initial cloud size $\sigma_{x0}$ in both Eqs.\,\ref{eq: Gaussian cloud contrast} and \ref{eq: Gaussian cloud deviation} emphasizes the importance of operation close to the point-source approximation $\sigma_{x0}\ll \sigma_f$, $k_x\sigma_{x0}\ll 1$.

The sensitivity of an interferometer to rotation, limited by projection noise, hinges on two additional factors: the number of atoms at each interferometer cycle ($N$) and the number of cycles per unit time ($\nu$). As far as the deviation $\Delta\Omega$ of Eq.\,\ref{eq: Gaussian cloud deviation} is neglected, this sensitivity ($\delta\Omega$) is given by the expression \cite{li_sensitivity_2024}:

\be \label{eq: PSI sensitivity}
\delta\Omega=\frac{\delta\Omega_{\rm single}}{\sqrt{\nu}}= \frac{T_{\rm{ex}}/T_R^2}{2C\sqrt{N\nu/2}K_{\text{eff}}\sigma_f}\,,
\ee
where it is preferable to use a longer interferometer duration ($T_R$) and enlarge the cloud final size ($\sigma_f$) to achieve good sensitivity.

Most Sagnac interferometers that serve as rotation sensors measure the Sagnac phase directly by probing the total atomic population or light intensity at an output port of a single loop. As the angular velocity of rotation is linearly proportional to the Sagnac phase, it follows that the smallest detectable rotation change is proportional to the smallest measurable phase change. Hence, sensitivity characterizes both the smallest detectable rotation and the uncertainty in detecting a given rate of rotation. By contrast, in the PSI, the minimum detectable rotation by the interferometer is determined by the condition that the cloud size exceeds the period of spatial oscillation, expressed as $k_x\sigma_f>1$. In addition, the contrast diminishes if the rotation rate exceeds the threshold where the fringe period becomes larger than the initial cloud size, namely $k_x\sigma_{x0}>1$. This implies a constraint on the fringe periodicity within the range $1/\sigma_f < k_x < 1/\sigma_{x0}$, leading to a limitation on the dynamic range of the interferometer, namely, $\Omega_{\rm min}<\Omega<\Omega_{\rm max}$, where from Eq.\,\ref{eq: PSI rotation phase} (and neglecting $\alpha$ from Eq.\,\ref{eq: Gaussian cloud output density}) we find

\be \label{eq: PSI limits}
\Omega_{\rm min}\sim \frac{T_{\rm{ex}}/T_R^2}{2K_{\text{eff}}\sigma_f}\,, \quad
\Omega_{\rm max}\sim \frac{T_{\rm{ex}}/T_R^2}{2K_{\text{eff}}\sigma_{x0}}\,,
\ee
such that $\Omega_{\rm max}/\Omega_{\rm min}\sim \sigma_f/\sigma_{x0}$. The exact value of the minimally detectable rotation $\Omega_{\rm min}$ depends on the specific method used for image analysis. In this context, previous works have discussed, for example, the ellipse fitting procedure \cite{foster_method_2002}, phase shear \cite{sugarbaker_enhanced_2013}, or phase map \cite{chen_robust_2020}. Finally, it should be noted that by combining Eqs.\,\ref{eq: PSI sensitivity} and \ref{eq: PSI limits}, the sensitivity in detecting a rotation rate, $\delta\Omega$, is associated with the minimally detectable rate, $\Omega_{\rm{min}}$, by the relationship $\delta\Omega / \Omega_{\rm{min}} \approx 1/\sqrt{N\nu/2}$. This suggests that for PSI, the sensitivity ($\delta\Omega$) solely pertains to uncertainty and may be much smaller (better) than the lowest detectable rotation rate.

\begin{figure}[ht!]
    \centering
    \includegraphics[trim = 40mm 0mm 40mm 15mm, clip, width=1\textwidth]{ 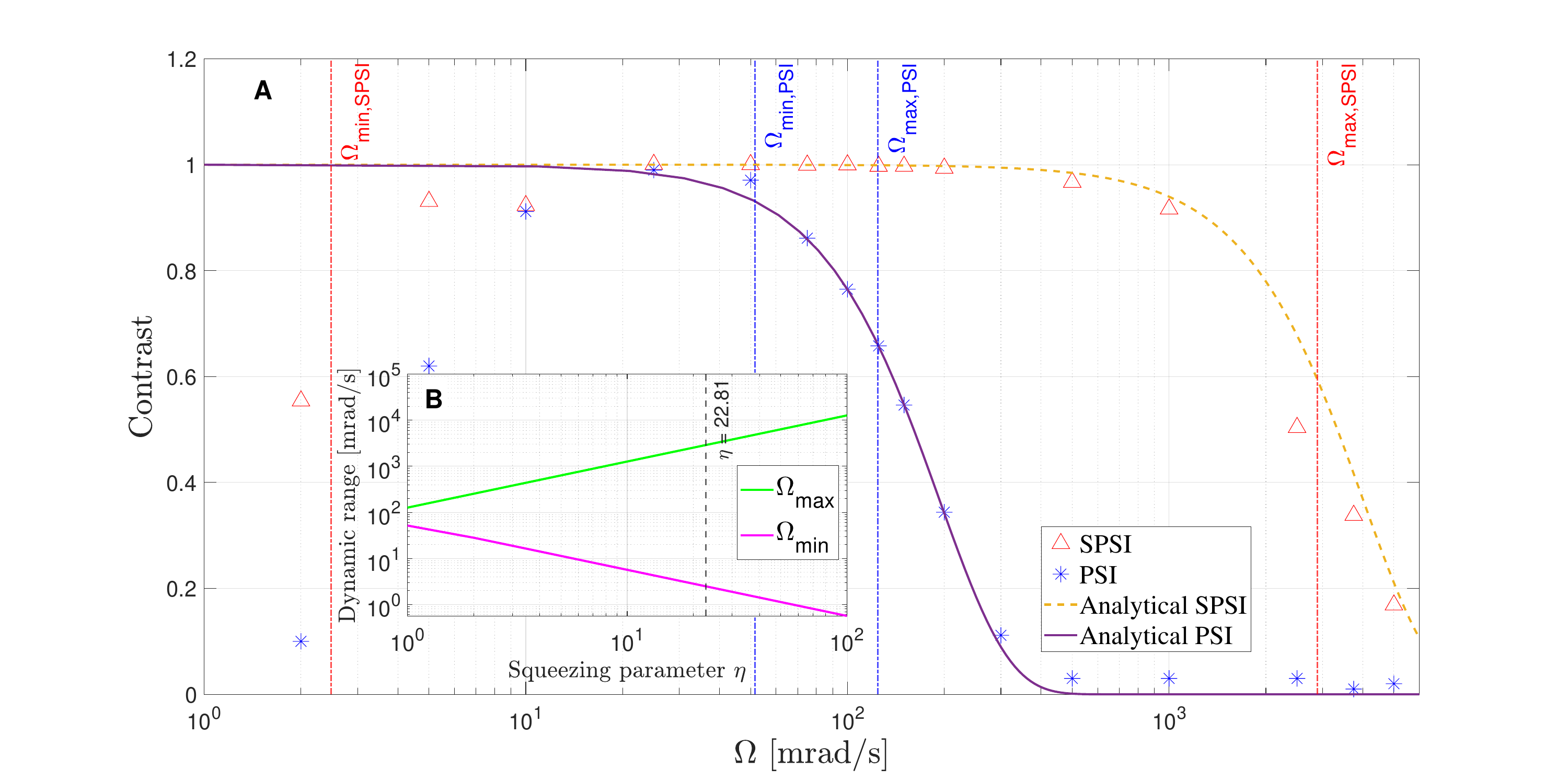}
    \caption{{\bf Contrast and dynamic range}. (A) Contrast vs. angular velocity ($\Omega$). Using $^{87}$Rb atoms, the simulation parameters are initial cloud size $\sigma_{x0} = 100\,\rm{\mu m}$, temperature $T = 5\,\rm{\mu K}$, and time between pulses $T_R = 5\,\rm ms$. For the repulsive potential, we take a beam power of $P=1\,$W with a cross-section of $400 \cross 400\,\mu$m$^2$ (according to the beam profile of Eq.\,\ref{eq:harmonic potential frequency}), blue-detuned by $\Delta = 2\pi\cdot10\,$GHz, and focused such that in the direction of acceleration it gives rise to an inverted harmonic potential and in the transverse direction a constant potential. The harmonic profile was optimized to achieve a large squeezing factor before the atoms exceed the region where the repulsive potential is harmonic. Following Eqs.\,\ref{eq: squeezing parameter} and \ref{eq:harmonic potential frequency}, and choosing $t_\text{acc} = 80.6\,\mu$s, we find a squeezing parameter of 22.81. The purple line represents the analytical solution of PSI contrast, while the dashed orange line corresponds to SPSI, according to Eq.\,\ref{eq: Gaussian cloud contrast}. The simulation points are the result of a numerical simulation. While the contrast reduction due to the ratio between the fringe periodicity and the initial cloud size determines the upper limit of the detection range, there exists a lower limit when the fringe period becomes larger than the final cloud size. These limits are roughly given by $\Omega_{\rm min}$ and $\Omega_{\rm max}$ (Eq.\,\ref{eq: PSI limits}), which are presented in the graph as vertical dashed lines. It is evident that using the SPSI greatly improves the contrast, thereby increasing the detection dynamic range. At high angular velocities, the contrast obtained from simulation decreases faster than the analytical solution due to the short spacing between fringes relative to the detection pixel size considered only in the numerical calculation. (B) The analytical curve of the detection range as a function of $\eta$ according to Eq.\,\ref{eq: PSI limits}. The dashed line presents the SPSI simulated in (A) with $\eta = 22.81$. }
    \label{fig:Contract and dynamic range}
\end{figure}

\subsection*{Squeezed Point Source Interferometer}

Integrating a repulsive potential that varies spatially to accelerate atom motion, prior to the interferometer sequence, has the capability to substantially enhance the operational efficiency and sensitivity of a PSI device.  Repulsive forces have been discussed in numerous contexts \cite{romero-isart_coherent_2017,yuce_quantum_2021,weiss_large_2021,ullinger_logarithmic_2022,neumeier_fast_2024,rozenman_observation_2024}, but here, the repulsive potential has the role of enlarging the area enclosed by the interferometer arms, while improving the crucial position-velocity correlation. This is achieved by applying a repulsive potential for an acceleration time $t_{\rm acc}$. To describe this position-momentum squeezing, let us define the coordinate system such that the $z$ axis is along the direction of the splitting and recombining laser beam. We may apply a quadratic repulsive potential along one or two axes transverse to $z$, but for the sake of simplicity, we describe only the dynamics along the $x$ coordinate. If the repulsive potential is quadratic $V_{\rm r}(x)=-(1/2) m\omega^2x^2$, where $m$ is the atomic mass, then after the acceleration stage, the initial coordinates in the direction $x$  of an atom are transformed as

\be \label{eq: harmonic pot. dynamics}
    \begin{aligned}
    x(t_{\rm acc}) &= \cosh(\omega t_{\rm acc})\cdot x(0)+\sinh(\omega t_{\rm acc})\cdot v_x(0)/\omega\,, \\ 
    v_x(t_{\rm acc}) &= \omega \sinh(\omega t_{\rm acc})\cdot x(0)+\cosh(\omega t_{\rm acc})\cdot v_x(0)\,,
    \end{aligned}
\ee 

It is easy to show that the phase-space distribution that forms after the repulsive pulse is exactly equivalent to the distribution that forms after free propagation for an effective duration $t_{\rm eff}$ when the initial distribution is squeezed with effective position uncertainty $\tilde{\sigma}_{x0}=\sigma_{x0}/\eta$ and velocity uncertainty $\tilde{\sigma}_{v0}=\eta\sigma_{v0}$, with the position-momentum phase-space squeezing parameter being

\be \label{eq: squeezing parameter}
\eta\equiv \sqrt{\cosh^2(\omega t_{\rm acc})+\frac{\omega^2\sigma_{x0}^2}{\sigma_{v0}^2}\sinh^2(\omega t_{\rm acc})}.
\ee

The effective time is given by

\be \label{eq: effective time}
t_{\rm eff}=\frac{\sinh(2\omega t_{\rm acc})}{2\omega\eta^2}\left(1+\frac{\omega^2\sigma_{x0}^2}{\sigma
_{v0}^2}\right)\,.
\ee

If $\omega t_{\rm acc}\ll 1$ and $r\equiv \omega \sigma_{x0}/\sigma_{v0}\ll1$ then $\eta\approx 1$ and $t_{\rm eff}\approx t_{\rm acc}$, such that the acceleration phase is ineffective. Conversely, if $\omega\gg \sigma_{v0}/\sigma_{x0}$ ($r\gg 1$) then even if the acceleration time is not long, such that $\omega t_{\rm acc}\lesssim 1$, then the squeezing parameter is large $\eta\sim r\omega t_{\rm acc}$ and the effective expansion time is $t_{\rm eff}\sim t_{\rm acc}/(\omega t_{\rm acc})^2$.  If the acceleration time is long, such that $\omega t_{\rm acc}>1$ the squeezing becomes exponentially large such that $\eta\approx \sqrt{1+r^2}e^{\omega t_{\rm acc}}$ and $t_{\rm eff}\approx 1/\omega$ is inversely proportional to the repulsive frequency. 
 
In a PSI, the final size of the cloud is limited not only by the duration of expansion but also by the extent of the Raman beam along the transverse direction. The intensity profile over the cloud size must be flat to ensure that the Rabi pulses are complete and do not introduce additional phase. In our analysis, we assume that a wide beam is used to ensure this requirement, although this might involve beam sizes of centimeters or a special beam-shaping procedure. In addition, we note that the velocity along the beam axis is limited by the requirement that velocity selection of the Raman beam will not cause inefficient transitions for a large part of atoms with the largest velocities. In our SPSI procedure, the atoms are accelerated along the transverse direction and hence this will not affect their interaction with the Raman beams unless the rotation frequency during the interferometer sequence is so large that transverse velocity components will have a considerable projection along the instantaneous longitudinal direction of the Raman beams.

The repulsive potential can be generated by a laser beam blue-detuned from the atomic resonance by $\Delta>0$. The effective ac Stark-shift is \cite{grimm_optical_2000}

\be \label{eq: potential stark shift}
V_{\rm ac}(\mathbf{r})=\hbar\frac{\Omega_R(\mathbf{r})^2}{4\Delta}=\frac{3\pi \Gamma}{2k_0^3 c}\frac{I(\mathbf{r})}{\Delta}\,,
\ee
where $\Omega_R$ is the local Rabi frequency, $k_0=2\pi/\lambda_0$ is the optical transition wave-vector, $\Gamma$ is the spontaneous emission rate, $I(\mathbf{r})$ is the light intensity, and $c$ is the speed of light. The inverse harmonic repulsive potential could be implemented by the quadratic intensity profile near the centre of a Gaussian beam propagating along the $z$ direction. However, a more efficient acceleration may be achieved by designing a fully quadratic beam shape. For simplicity, we consider here a one-dimensional model of acceleration in the $x$ direction induced by a laser beam propagating along the $y$ direction and having a homogeneous profile in the $z$ direction in the volume containing the atom cloud. The light beam profile is  $I(x,z) = I_0(1-4 x^2 / \Delta x_\text{rep}^2)$ for $|x| < \Delta x_\text{rep} /2 $, $|z| < \Delta z_\text{rep} / 2$ and zero otherwise. Here the peak intensity $I_0$ is related to the beam power $P$ as $I_0=3P/2\Delta x_\text{rep}\Delta z_\text{rep}$. The harmonic frequency of the repulsive potential is then

\be \label{eq:harmonic potential frequency}
\omega_\text{Harmonic} = \sqrt{\frac{18\pi \Gamma P}{ m k_0^3 c \Delta \cdot \Delta x_\text{rep}^3 \Delta z_\text{rep}}}.
\ee

The harmonic profile is quite advantageous for achieving a large squeezing factor before the atoms reach the region with a repulsive potential. For example, for the parameters introduced in Fig.\,\ref{fig:Contract and dynamic range}, $\omega_\text{Harmonic} = 2\pi \cdot 1220\,$Hz.

The optical repulsive potential has the consequent effect of heating the atoms due to scattered photons \cite{grimm_optical_2000}. Along the direction of beam propagation, the atoms gain momentum in correlation to the spatial intensity $I(\textbf{r})$, which can be compensated by shifting the initial cloud position or, if necessary, by employing a counter-propagating beam. In addition, the scattering induces a random walk in velocity space in all directions, potentially enlarging the effective initial cloud size, $\tilde{\sigma}_{x0}$, and consequently decreasing the upper detection limit $\Omega_{\text{max}}$ (Eq.\,\ref{eq: PSI limits}). Under the limitations we have taken for the laser power and typical parameter values, this effect reduces $\Omega_{\text{max}}$, e.g., by $\sqrt{2}$ for the parameters of Fig.\,\ref{fig:Contract and dynamic range}. This effect can be reduced to a negligible level by improving the beam parameters, for example, by equally increasing the beam power $P$ and the detuning $\Delta$. More details about the heating process can be found in the supplementary information. One could also explore using different types of potential, such as magnetic gradients, to mitigate heating.

Let us note that effective phase-space squeezing is also the natural result of the expansion of a Bose-Einstein condensate (BEC) with repulsive atom-atom interactions. A BEC that is initially trapped in a cigar-shaped potential expands along its fast axes, after turning off the trap, such that its size scales like $\sigma(t)\approx \sigma(0)\sqrt{1+\omega^2 t^2}$, where $\omega$ is the trap frequency\,\cite{castin_bose-einstein_1996}. After a time $t\gg \omega^{-1}$ the momentum width of the expanding wave packet $\sigma_p\approx m\sigma(0)\omega$     corresponds to an equivalent wave packet with no atom-atom interactionswith an  initial position uncertainty $\sigma_{x0}=\hbar/2\sigma_p$. This initial equivalent wave-packet size is smaller than the minimum uncertainty wave packet with $\sigma_{\rm min}=\sqrt{\hbar/2m\omega}$ by a factor $\eta_{BEC}=\sigma_{\rm min}/\sigma(0)< 1$ and smaller than the actual initial BEC wave-packet by a factor $\eta_{BEC}^2$. This property may be used to create a collimated beam by the method of delta-kick cooling, but in our case it may be an advantage  without this kick due to the fast expansion and high velocity-position correlation. However, in this paper we concentrate on an initial thermal atomic cloud without atom-atom interaction due to its practical advantages over a BEC in this context. 

\begin{figure}[ht!]
    \centering
    \includegraphics[trim = 30mm 0mm 40mm 15mm, clip, width=1\textwidth]{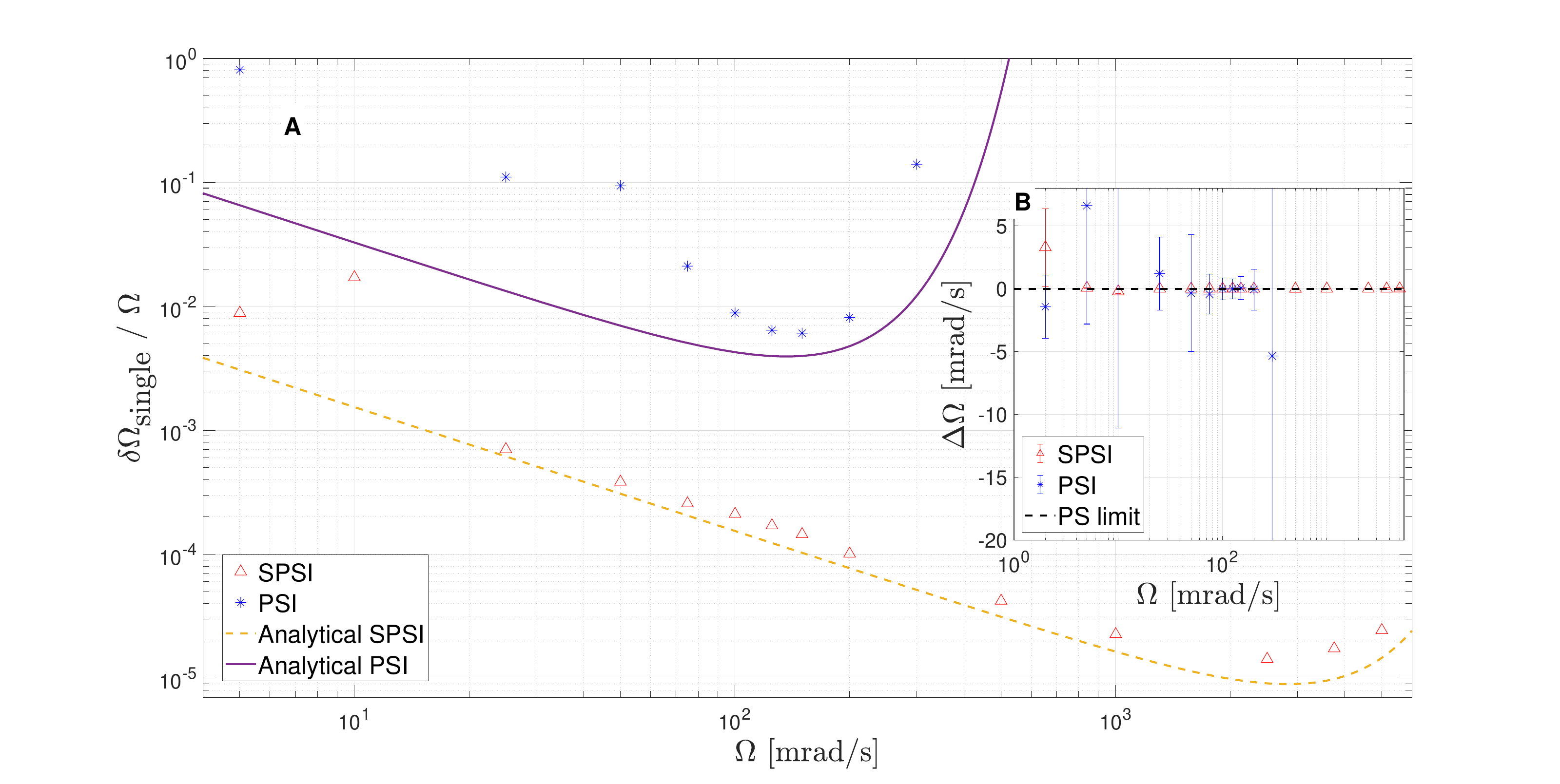}
    \caption{{\bf Relative single-shot sensitivity and measured angular velocity deviation}. (A) Relative sensitivity $\delta \Omega_\text{single} / \Omega$ vs $\Omega$, where $\Omega$ is the nominal angular velocity, for the same simulation parameters and notation as in Fig.\,\ref{fig:Contract and dynamic range}. The analytical curves are determined according to Eq.\,\ref{eq: PSI sensitivity}. The simulation results for PSI at large angular velocities where $\delta\Omega_\text{single}/\Omega>1$ are irrelevant and therefore not shown. (B) Simulation results of the measured angular velocity deviation $\Delta \Omega$ (Eq.\,\ref{eq: Gaussian cloud deviation}) for different angular velocities. Here $\Delta \Omega$ the discrepancy between the measured and nominal angular velocities, while the sensitivity (uncertainty) $\delta\Omega_\text{single}$ is represented by the error bars. $\Omega$ is extracted from each simulation run and obtained from the fringe spatial frequency of the fitted sinusoidal pattern (as described in Fig.\,\ref{fig:Simulation Example}). The data point labels are the same as in (A), while the black dashed line illustrates the point-source (PS) limit of zero deviation. The simulation results for PSI are not shown for large angular velocities where $\delta\Omega_{\rm single} / \Omega > 1$. The enhanced performance of the SPSI in dynamic range and sensitivity (measurement uncertainty) is clearly visible.}
    \label{fig:Relative-sensitivity and Delta omega}
\end{figure}

\subsection*{Performance Analysis}

To estimate the performance of the SPSI and compare it to that of the standard PSI, we calculated its characteristics analytically and numerically (see Materials and Methods for the latter).

In Fig.\,\ref{fig:Contract and dynamic range}(A), the fringe pattern's contrast is plotted against the angular velocity for both methods. This illustrates that SPSI can measure higher angular velocities as its contrast diminishes at higher angular velocities, much larger than for standard PSI. We briefly note that at large angular velocities, the fringe spatial frequency becomes high, and the detection pixel resolution becomes a limiting factor, causing the numerical contrast to decay more rapidly than the analytically predicted decay, which does not account for this limitation. Similarly, the SPSI can measure lower angular velocities than the PSI because its final cloud radius is larger, rendering it more sensitive to slow angular velocities characterized by low spatial fringe frequency. This improvement in both limits is depicted in Fig.\,\ref{fig:Contract and dynamic range}(B) by the analytical curve of the detection range as a function of $\eta$ according to Eq.\,\ref{eq: PSI limits}.

Fig.\,\ref{fig:Relative-sensitivity and Delta omega}(A) presents a comparison of the relative sensitivity $\delta \Omega_\text{single} / \Omega$ between the two methods. It demonstrates an improvement of over one order of magnitude when both methods operate within their dynamic range $(60\lesssim \Omega \lesssim 200\,\rm{mrad/s})$, with a notably superior ratio beyond that range. Additionally, in Fig.\,\ref{fig:Relative-sensitivity and Delta omega}(B), the plot of angular velocity deviation $\Delta \Omega$ (Eq.\,\ref{eq: Gaussian cloud deviation}) provides further support for the advantage of SPSI.

Having established the clear advantage of the SPSI when compared to PSI with equivalent interferometer durations, where the transverse size of the final cloud is much larger for the SPSI, we examine the SPSI's advantages in terms of spatial compactness, particularly in scenarios with reduced operational timeframes where a conventional PSI would be nonfunctional due to an inadequate final cloud size relative to the initial cloud dimensions. To quantify the SPSI's interferometric compactness, we introduce the dimensionless parameter $a_t = T_{R,SPSI} / T_{R,PSI}$, which represents the Ramsey time reduction in a compact SPSI compared to a standard PSI ($a_t < 1$). This reduction would also reduce the fall distance $h$ during the interferometer by $a_t^2$ and the repetition rate $\nu$ by $a_t$. Additionally, the radius of the final cloud scales as $\sigma_f \sim a_t T_{\rm{ex}} \sigma_{v0} \eta$. Consequently, the sensitivity given by:

\be \label{eq: SPSI sensitivity}
\delta\Omega \approx \frac{1}{a_t T_R^2\cdot 2K_{\text{eff}}C\sqrt{N \nu/2 a_t}\cdot a_t \sigma_{v0} \eta},
\ee
such that the sensitivity per shot ratio become $\delta\Omega_{SPSI}/\delta\Omega_{PSI}=(C_{PSI}/C_{SPSI})/ (a_t^{1.5} \cdot \eta$), where it was assumed that the interferometer starts right after the beginning of the expansion, such that $T_{\rm ex}\approx 2T_R$. For the design of a rotation sensor, it would be necessary to compromise between sensitivity and compactness and to identify the optimal operational point based on the parameters $\eta$ and $a_t$. Fig.\,\ref{fig:Sensativity(eta,a_t)} illustrates the advantage in sensitivity offered by the SPSI over the standard PSI for different parameter values. It is evident from the figure that the incorporation of a repulsive potential allows for a sensitivity enhancement of up to two orders of magnitude while concurrently reducing the cycle time by a factor of approximately two, or alternatively, decreasing the cycle time by around tenfold without compromising sensitivity.

\begin{figure}[ht!]
    \centering
    \includegraphics[trim = 25mm 5mm 20mm 15mm, clip, width=1\textwidth]{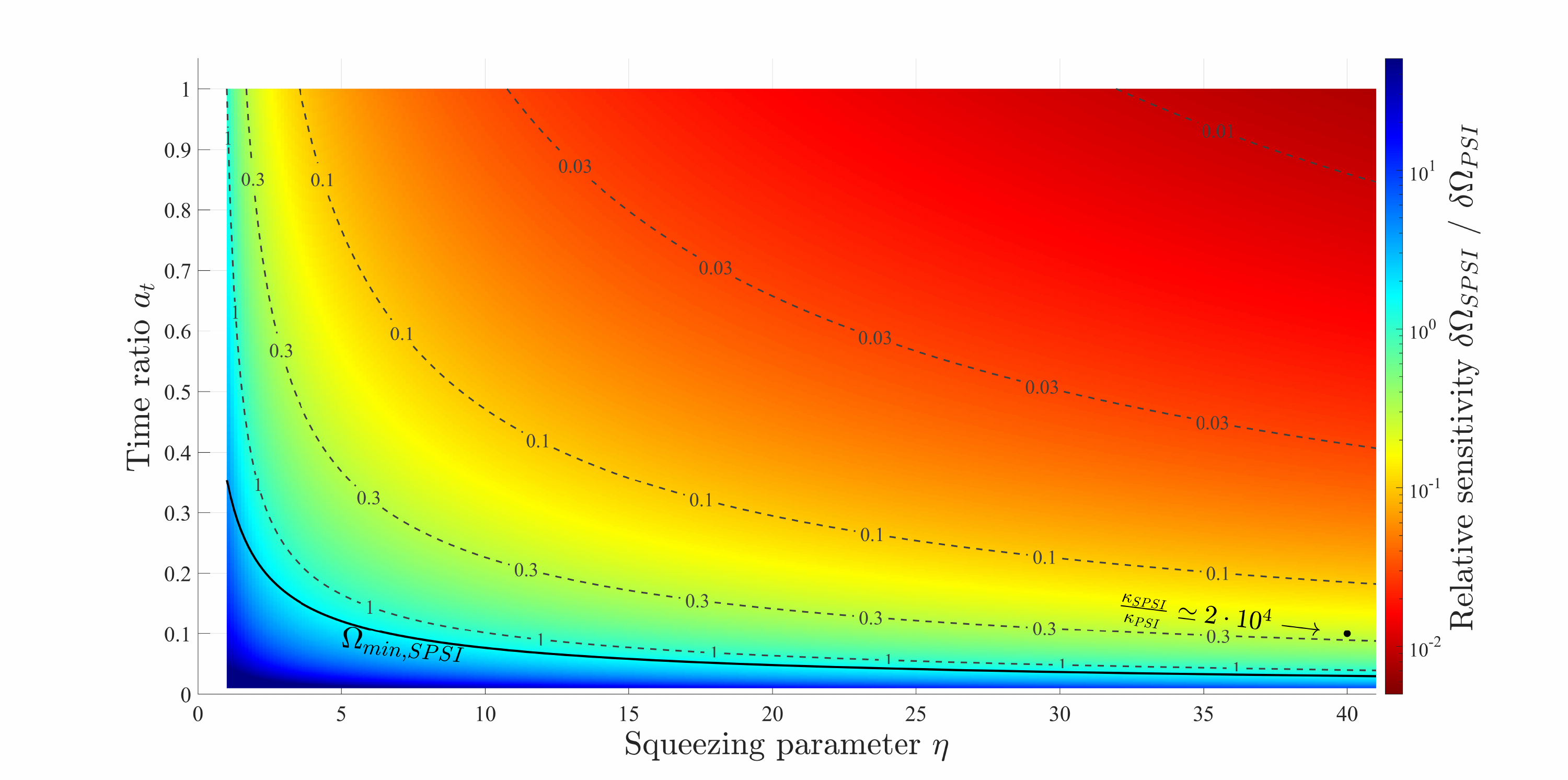}
    \caption{{\bf Sensitivity-compactness trade-off in the SPSI}. The sensitivity ratio $\delta\Omega_{SPSI}/\delta\Omega_{PSI}$ based on Eq.\,\ref{eq: SPSI sensitivity}, is colour coded against the squeezing parameter $\eta$ and the cycle time ratio $a_t = T_{SPSI} / T_{PSI}$ for an angular velocity of $\Omega = 100\,\rm{mrad/s}$, while maintaining constant values for the remaining parameters: Time between pulses $T_R = 10\,\rm{ms}$, temperature $T = 2\,\rm{\mu K}$, and initial cloud size $\sigma_{x0} = 100\,\rm{\mu m}$. The maximum value of $\eta$ was chosen to ensure that the majority of atoms are affected by the same potential throughout the acceleration phase. The bold black line indicates the validity boundary of this plot beyond which the minimal detectable angular velocity of the SPSI method (Eq.\,\ref{eq: dynamic range - short time}) exceeds the nominal angular velocity $\Omega$. It is evident that the implementation of a repulsive potential allows for a sensitivity enhancement of more than an order of magnitude while simultaneously reducing the cycle time by a factor of about two (as indicated by the bright red colour) or reducing the cycle time by approximately tenfold without affecting the sensitivity (as indicated by the light-blue colour). Specifically, utilizing the compactness factor definition of Eq.\,\ref{eq: compactness factor}, we find that with $\eta=40$ and $a_t=0.1$, we have a performance enhancement of $\kappa_{SPSI}/\kappa_{PSI} \simeq 2\cdot 10^4$ (black point).}
    \label{fig:Sensativity(eta,a_t)}
\end{figure}

\subsection*{Chip-Scale Device}

Let us now compare between standard PSI and SPSI within the constraints of identical Ramsey times ($a_t=1$) and fall heights $h$ in the context of chip-scale devices. Employing the repulsive potential to expand acceleration parallel to the chip's plane enables the SPSI method to optimize the device's confinement.

 For the comparison, under the same initial conditions (temperature, cloud size, and number of atoms), the ratio between the sensitivities obtained in the two methods is $\delta\Omega_{SPSI} / \delta\Omega_{PSI} \approx (C_{PSI}/C_{SPSI}) / \eta$. Within the dynamic range of both methods, the contrast ratio is approximately $\sim 1$, while the detection limits ratio is $\Omega_{\rm min, SPSI} / \Omega_{\rm min, PSI} = 1 / \eta$ and $\Omega_{\rm max, SPSI} / \Omega_{\rm max, PSI} = \eta$. Consequently, for chip-scale devices, we can expect the sensitivity improvement of about $\eta$ and an increase in the dynamic range $\Omega_{\rm max} / \Omega_{\rm min}$ by a factor of $\eta^2$.

Specifically, we consider the following chip-scale scenario: a rectangular vacuum cell with dimensions of $1\,\rm{mm} \times 10\,\rm{mm} \times 10\,\rm{mm}$, with the short dimension aligned perpendicular to the direction of the repulsive potential's acceleration. The SPSI sequence commences with the capture of a thermal atomic cloud comprising $N = 10^6$ $^{87}$Rb atoms, with a radius of $\sigma_{x0} = 100\,\rm{\mu m}$ and temperature of $T = 2\,\rm{\mu K}$, using a magneto-optical trap and optical molasses setup. Employing a fully harmonic potential characterized by dimensions of $\Delta x_\text{rep} = 400\,\rm{\mu m}$ and $\Delta z_\text{rep} = 400\,\rm{\mu m}$, with total power of $P = 1\,\rm{W}$ and an acceleration time of $t_{\rm acc} = 0.1\,\rm{ms}$, yields a squeezing parameter $\eta = 46.2$, potential frequency $\omega_\text{Harmonic} = 2\pi \cdot 1220\,$Hz, and final velocity uncertainty $\tilde{\sigma}_{v0} = 0.64\,$m/s. Following acceleration, the atoms undergo a Ramsey sequence with a time $T_R = 2.5\,\rm{ms}$, resulting in a total fall height of $h = 120\,\rm{\mu m}$. The final cloud dimensions are  $\sigma_{f,z} = 170\,\rm{\mu m}$ height and $\sigma_{f,x} = 0.33\,\rm{cm}$ width. Without heating, a vacuum dimension of $0.5\,\rm{mm}$ would suffice, but if heating is not mitigated, a vacuum dimension of $1\,\rm{mm}$ would be required, or alternatively, one would have to take into account that a considerable amount of atoms could not be recaptured for the next cycle, and reloading atoms from a source would be required.

With a number of operations per second $\nu = 100\,\rm{s^{-1}}$, the SPSI exhibits a sensitivity of $\delta \Omega = 1\,\rm{\mu \rm{rad}/(s\cdot \sqrt{Hz})}$ and a one-shot sensitivity of $\delta \Omega = 10\,\rm{\mu\rm{rad}/s}$. The minimum and maximum detectable angular velocities are $\Omega_{\rm min} = 7.5\,\rm{mrad/s}$ and $\Omega_{\rm max} = 11.47\,\rm{rad/s}$, respectively.

Studying the impact of heating due to the repulsive optical potential in a chip-scale device is crucial. The atoms acquire momentum in the direction of the optical repulsive potential beam, and this effect can be managed, as discussed earlier. However, heating in the direction of the splitting and recombining laser beam presents more challenges, especially when we aim for the smallest possible chip. For the repulsive potential parameters outlined above, the calculation of the size of the cloud along the direction of splitting results in $\sigma_{z,f} \approx 300\,\mu\rm{m}$, such that most of the atoms remain within the $1\,\rm{mm}$ of the device. Additionally, as mentioned earlier, the dynamic range is restricted due to heating and is reduced in this example by a factor of approximately 3.

In a fully integrated chip-scale device it will be necessary to design photonics that will provide the shapes of the laser beams responsible for the repulsive potential and the stimulated Raman transitions such that the former has the quadratic intensity profile along $x$ and the latter has a flat intensity profile over the atomic cloud in both $x$ and $y$ directions. A beam that has a flat intensity across the significant spatial extent of the atomic cloud in $x$, can be generated, for example, using a set of beam splitters, as commonly implemented in 2D-MOT configurations \cite{chaudhuri_realization_2006}. 

\section*{Discussion}

The presented method offers multiple orders of magnitude of improvement in the sensitivity and dynamic range of point-source interferometers. Our theoretical and numerical analysis demonstrates the potential of SPSI to significantly enhance performance in both standard-sized and chip-scale devices. The results highlight the substantial benefits of position-momentum phase-space squeezing, which include shorter cycle times and higher repetition rates. While our findings suggest that SPSI can pave the way for new practical applications, it is essential to further investigate the more minute and technical details of the method. Future experimental implementations should focus on optimizing parameters such as the repulsive potential, cloud size, Raman beam's profile and detection techniques to fully realize the theoretical advantages of SPSI.

Analysis of the simulation results from Fig.\,\ref{fig:Sensativity(eta,a_t)} reveals that employing squeezing can lead to a tenfold reduction in the cycle time without compromising sensitivity. Conversely, sensitivity enhancement by up to two orders of magnitude is achievable while concurrently reducing the cycle time by about two-fold. 

The ratios of the detection limits of PSI and SPSI are determined by:

\be \label{eq: dynamic range - short time}
\frac{\Omega_{\rm min, SPSI}}{\Omega_{\rm min, PSI}} = \frac{1}{a_t^2 \cdot \eta}, \quad
\frac{\Omega_{\rm max, SPSI}}{\Omega_{\rm max, PSI}} = \frac{\eta}{a_t}.
\ee

Overall, the ratio $\Omega_{\rm max} / \Omega_{\rm min}$ changes by a factor of $\eta^2 \cdot a_t$. Consequently, in the case of a tenfold reduction in the cycle time, the dynamic range would increase by more than two orders of magnitude.

To better quantify the performance improvement, we introduce a compactness factor

\be \label{eq: compactness factor}
\kappa \equiv\frac{\Omega_{\text{max}}/\Omega_{\text{min}}}{\delta\Omega\cdot h}.
\ee

This factor encapsulates sensitivity $\delta \Omega$, dynamic range (as the ratio $\Omega_{\rm max}/\Omega_{\rm min}$, and the vertical size of the interferometer $h$. Based on the previous comparison between SPSI and PSI, we observe an improvement ratio of $\eta^3 \cdot \sqrt{a_t}$, which can exceed four orders of magnitude. For instance, considering the conditions depicted in Fig.\,\ref{fig:Sensativity(eta,a_t)}, with $\eta=40$ and $a_t = 0.1$, we derive $\kappa_{\rm SPSI} / \kappa_{\rm PSI} \simeq 2 \cdot 10^4$.

Finally, we note that details of the numerical analysis are presented in Materials and Methods, while the AC Stark interaction, heating by light and a full quantum approach are discussed in the supplementary information. As an outlook, let us note that accelerating the cloud expansion may also be realized by other interactions, such as repulsive magnetic forces or attractive electric forces. An accelerated expansion can even be acquired passively by diffracting a beam of atoms (e.g., from a 2D MOT) off a slit. An interesting problem left for future work is to find the optimal functional form of the repulsive potential.

To conclude, the SPSI described in this work enables either enhanced performance for standard-size devices or maintaining the performance while miniaturizing it to a chip-scale device, opening the door to real-life applications.

\begin{figure}[ht!]
    \centering
    \includegraphics[trim = 20mm 5mm 20mm 10mm, clip, width=1\textwidth]{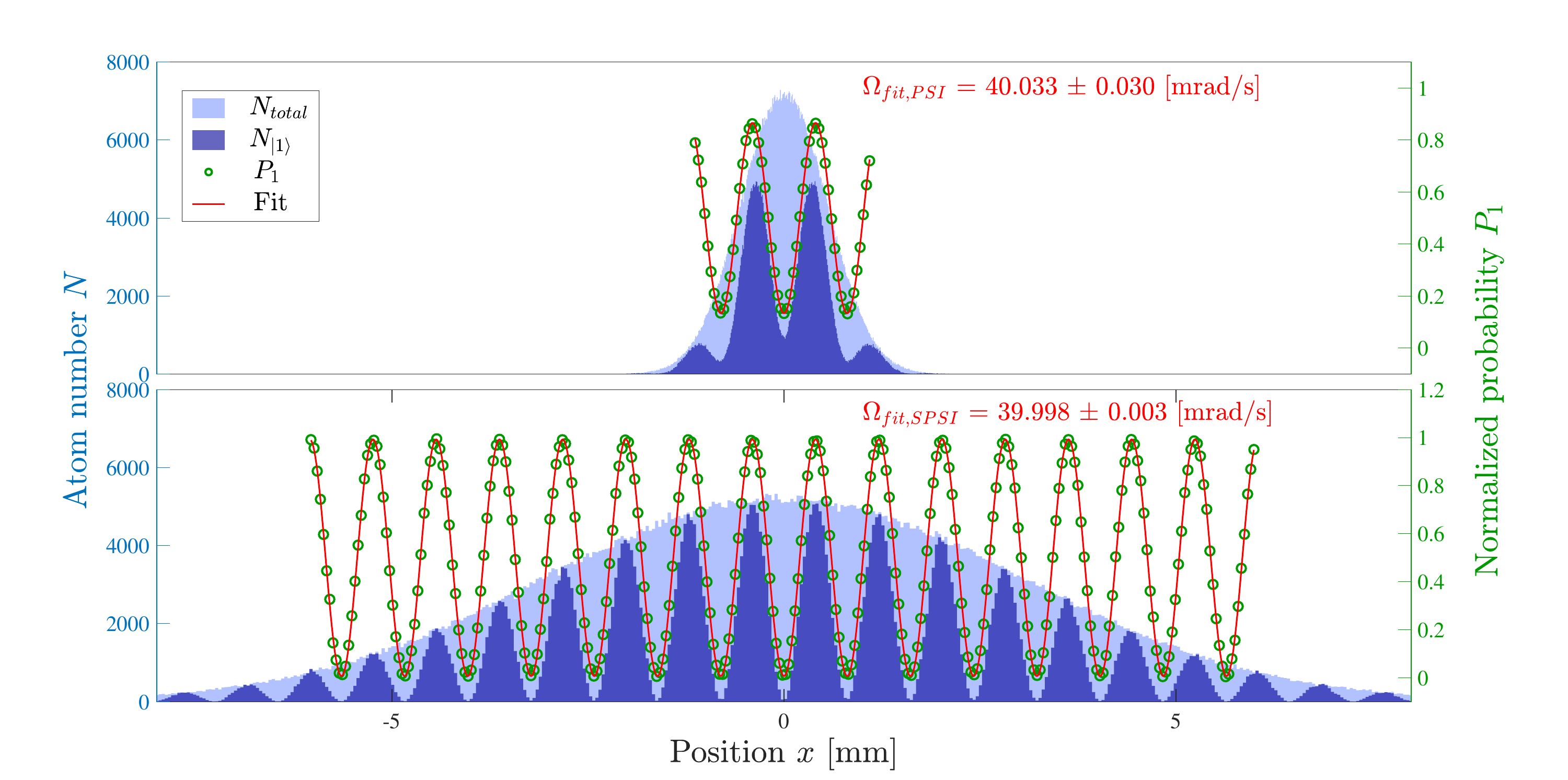}
    \caption{{\bf Simulation fringe pattern fit}. This figure compares the simulation results of both methods, the top for PSI and the bottom for SPSI. The light blue shaded area depicts the total number of atoms in each bin, denoted as $N_{\text{total}}$, while the darker shade indicates only those atoms in the internal state $|1\rangle$, denoted as $N_{|1\rangle}$. The green points represent the normalized probability for the $|1\rangle$ state, calculated as $P_1 = N_{|1\rangle}(x)/N_{\text{total}}(x)$, while the red line denotes the sinusoidal fit to these green points. The angular velocity is determined from the fit's fringe spatial frequency using Eq.\,\ref{eq: Gaussian cloud output density}. In this example the initial 1D Gaussian cloud comprising $N=10^6$ atoms, with a width of $\sigma_{x0}=100\,\rm{\mu m}$ and a temperature of $T=5\,\rm{\mu K}$, as in Figs.\,\ref{fig:Contract and dynamic range} and \ref{fig:Relative-sensitivity and Delta omega}. For the PSI method's simulation, the atoms undergo free expansion during the interferometer sequence, with a time interval between pulses of $T_R = 25\,\rm ms$ used in this figure, while the simulated angular velocity is $\Omega = 40\,\rm{mrad/s}$. In the case of SPSI, the cloud is additionally subjected to a repulsive potential having a power of $P = 1\,\rm{W}$, with a beam cross-section of $400 \cross 400\,\mu$m$^2$, as in Figs.\,\ref{fig:Contract and dynamic range} and \ref{fig:Relative-sensitivity and Delta omega}, activated for $t_{\text{acc}} = 20\,\rm{\mu s}$. This results in a potential frequency of $\omega = 2\pi\times 1.214\,\rm{kHz}$ (Eq.\,\ref{eq:harmonic potential frequency}), a squeezing parameter $\eta = 5.4$ (Eq.\,\ref{eq: squeezing parameter}), and an effective time $t_{\text{eff}} = 0.84\,\rm{ms}$ (Eq.\,\ref{eq: effective time}). Consequently, the cloud size increases notably, and we chose the potential parameters here such that it would be easy to compare to the PSI simulation. Note that the contrast and number of fringes in the PSI method is smaller than that of the SPSI (see Figs.\,\ref{fig:Phase-Space Squeezing}-\ref{fig:Contract and dynamic range}), leading to a fit deviation and uncertainty one order of magnitude higher for the PSI case. It is important to note that while the histograms appear different in scale, the total atom count remains constant for both methods. The discrepancy in appearance is due to different bin sizes used in the PSI and SPSI plots to accommodate the varying cloud expansions, with SPSI utilizing larger bins to cover its greater spatial extent.}
    \label{fig:Simulation Example}
\end{figure}

\section*{Materials and Methods}
\subsection*{Numerical Analysis}

To examine the impact of the SPSI, we conduct detailed numerical simulations for PSI and SPSI. Initially, we generate a cloud of $N$ atoms following the cooling phase, characterized by a spatial Gaussian distribution with a standard deviation of $\sigma_{x0}$. As we deal exclusively with thermal clouds, the atoms' velocities follow a Maxwell-Boltzmann distribution corresponding to the given temperature. In a standard PSI simulation, the cloud undergoes expansion, followed by the initiation of the atomic-interferometer sequence with a time interval $T_R$ between pulses. Upon completion of this sequence, we determine each atom's internal state via Eq.\,\ref{eq: state probability}, utilizing a Monte Carlo method. The wave-vector $k_v$ and the phase $\phi_s$ are determined by the rotation angular velocity (parameter $\Omega$) as in Eq.\,\ref{eq: state probability}. We simulate absorption imaging of the two atomic states across the cloud by counting the atoms of each state in bins of spatial locations and normalizing the counts of one state to the total population in each bin $P_1(x) = N_{|1\rangle}(x)/N_{\rm total}(x)$. This approach reveals the interferometer's fringe pattern as per Eq.\,\ref{eq: PSI rotation phase}, illustrated in Fig.\,\ref{fig:Simulation Example}. Utilizing a sinusoidal fit allows us to extract the fringe's periodicity, $k_x$, facilitating the calculation of angular velocity based on Eq.\,\ref{eq: Gaussian cloud output density}. The $\alpha$ factor is accounted for by scaling the fit's frequency accordingly before extracting the angular velocity as described in Eq.\,\ref{eq: Gaussian cloud output density}. The uncertainty in angular velocity, denoted as single-shot sensitivity $\delta \Omega_\text{single}$, is computed based on the fit uncertainty. Running simulations for multiple repetitions and calculating the standard deviation provides a comparable result to the uncertainty derived from a single run's fit.

In an SPSI simulation, we incorporate an additional stage: accelerating the atomic ensemble via a repulsive potential. We implement a blue-detuned inverted harmonic-shaped optical potential (Eq.\,\ref{eq:harmonic potential frequency}) characterized by a total power $P$ and beam dimensions $\Delta x_{rep}$ and $\Delta z_{rep}$, applied over a brief duration of $t_{\rm acc}$. By numerically solving the atomic dynamical system using an ordinary differential equation (ODE) solver, we quantify the acceleration's kinematic effects. Upon completion of the acceleration stage, the simulation protocol proceeds consistently with a standard PSI methodology.

The normalized population is fitted to a sinusoidal function, $f(x) = A \sin (Bx+C) +D$, which initially employs an FFT algorithm to identify optimal frequencies and uses them as initial guesses for the fit algorithm based on non-linear least-squares. To improve the fit's success rate, we implement a basic image processing stage before normalizing $P_1$. This stage involves applying different averaging windows to the image's pixels to reduce noise in the atom count in each bin. Additionally, the algorithm examines various windows of interest from the image, excluding the noisier edges where fewer atoms are counted and more fluctuations occur.

\textbf{Acknowledgments}\\
This work was partly supported by the Israel Science Foundation Grants No. 858/18, 1314/19, 3515/20 and 3470/21.

\bigskip

\appendix 

\section*{Supplementary information}

\subsection*{Repulsive potential by AC Stark shift}

In this section, we provide more details about the preparation of the atoms and the repulsive potential and discuss the effects of heating induced by the light beam responsible for the repulsive potential. The preparation sequence for the atomic cloud begins with a 3D magneto-optical trap, followed by optical molasses to reduce the cloud's temperature. Before applying the repulsive potential, the $^{87}$Rb atoms are optically pumped to the $|F=2,m_F=2\rangle$ atomic state. The repulsive potential is blue-detuned with circularly polarized $\sigma^+$ light, which ensures that any undesired excitation of the atoms decays back to the $|F=2,m_F=2\rangle$ state. The repulsive potential defines the quantization axis, eliminating the need for external magnetic fields. At the end of the acceleration and just before the interferometer sequence, the atoms are pumped to the non-magnetic $|F=1,m_F=0\rangle$ state so that the interferometric sequence may begin.

The blue-detuned light induces a positive-energy shift of the ground atomic level and a negative-energy shift of the excited level. As the energy shift is proportional to the light intensity, an atom in the ground level is repelled from the high-intensity beam centre while the excited level is drawn towards it.  For the repulsion to be effective and to avoid heating by spontaneous emission, it is essential that $\Omega_R^2 \ll \Delta^2$ so that the excited-level occupation stays small. For example, for the repulsive potential parameters used in Fig.\,2, in the centre of the potential ($I(x,z)=I_0$) we get $\Omega_R = 2\pi \cdot 3.16\,\rm{GHz}$, such that $\Delta = 2\pi\cdot10\,\rm{GHz}$ gives a ratio of $\Delta^2 / \Omega_R^2 = 10$.

An unavoidable effect of the repulsive potential is the heating of the atoms induced by scattered photons\cite{grimm_optical_2000}. In the longitudinal direction of propagation of the light beam, the atoms experience an average heating of $4E_\text{rec}/3$ per scattering event, while in the two transverse directions, the heating amounts to $E_\text{rec}/3$ per scattering process, where $E_\text{rec}$ denotes the recoil energy, determining the recoil velocity $v_\text{rec} = \sqrt{2E_\text{rec} / m}$. Thus, the velocity increment in each axis for every scattering event equals $v_\text{rec}/\sqrt{3}$. The heating process along the acceleration direction contributes to an enlargement of the effective initial cloud's spatial width, denoted as $\tilde \sigma_{x0,\text{eff}} = \sqrt{(\sigma_{x0}/\eta)^2 + (t_\text{eff} \cdot \sigma_{v0, \text{heat}})^2}$, where $\sigma_{v0, \text{heat}} = \sqrt{N_{\text{sc}} / 3}\cdot v_\text{rec}$ represents the heating due to the scattered photons from the repulsive potential. Here, $N_\text{sc} = \Gamma_{\text{sc}} t_\text{acc}$ signifies the number of scattering events, with $\Gamma_\text{sc} = (3\pi \Gamma^2 I(\textbf{r}))/ (2\hbar k_0^3 c \Delta^2)$ denoting the scattering rate. In the SPSI method, the effective reduction of $\sigma_{x0}$ augments $\Omega_{\text{max}}$, and this heating mechanism solely diminishes its value without compromising the sensitivity within the new detection range. If $(\sigma_{x0}/\eta)^2 \gg (t_\text{eff} \cdot \sigma_{v0, \text{heat}})^2$, the effect of the scattering on the phase-space distribution becomes negligible. For the repulsive potential parameters used in Fig.\,2, the relation is $(\sigma_{x0} / \eta) ^ 2 / (t_\text{eff} \cdot \sigma_{v0, \text{heat}})^2 \approx 1$, and therefore in this example, heating only decreases $\Omega_\text{max}$ by a factor of $\approx \sqrt2$. A more detailed derivation will be provided in the following section.

The repulsive potential force in the direction of optical beam propagation adds a velocity in this axis equal to $v_\text{long} = N_\text{sc} v_{\text{rec}}$. Using the same potential and interferometer parameters as in Fig.\,2, $v_\text{long}$ is equal to $28\,\rm{cm/s}$, resulting in a total longitudinal displacement of $\Delta x_\text{long} = 2.8\,\rm{mm}$. This can be easily addressed by enlarging the vacuum cell or by setting the initial cloud away from the centre of the cell.

\subsection*{Free evolution of a Gaussian phase-space distribution and the effect of heating}

In this section we describe the evolution of a Gaussian position-momentum phase-space distribution due to free propagation, which is valid for both classical and quantum systems, and discuss the effect of heating on such a distribution. In a two-dimensional phase space of position $x$ and velocity $v$, a Gaussian distribution in the form of an exponent of a bi-quadratic expression in $x$ and $v$ remains Gaussian in either free evolution or evolution in a quadratic potential. This conservation of the Gaussian form follows from the fact that any linear transformation of a bi-quadratic expression of two variables remains bi-quadratic under such a transformation. Any Gaussian form where the distribution is centred around $x=0$ and $v=0$ can be written in the form

\be 
\rho(x,v,t)\propto e^{-(x-vt)^2/2\sigma_{x0}^2}e^{-v^2/2\sigma_{v0}^2}\,, 
\label{eq:uncorrelated} \ee
which is a distribution that evolves under free propagation from an uncorrelated initial distribution at $t=0$ with a spatial uncertainty $\sigma_{x0}$ and velocity uncertainty $\sigma_{v0}$. This implies that any Gaussian distribution in phase space can be obtained by free propagation over an effective time $t_{\rm eff}$  starting from an uncorrelated distribution.  This is the way that the effective squeezed distribution with a squeezing factor $\eta$ is obtained from the distribution that evolves under the repulsive inverse harmonic potential after acceleration over a time $t_{\rm acc}$. It is easy to acknowledge these properties by noting that a Gaussian distribution in phase space has an elliptical shape in the $x-v$ plane, and any ellipse can be transformed into an ellipse whose axes are aligned along the main axes by a proper rotation transformation. 

The quadratic expression in the exponent can also be written in an alternative form that emphasizes the position-velocity correlations in the distribution: 

\be  \frac{(x-vt)^2}{2\sigma_{x0}^2}+\frac{v^2}{2\sigma_{v0}^2}=\frac{x^2}{2\sigma_f^2}+\frac{(v- x/T)^2}{2\sigma_{vt}^2}\,, 
\label{eq:quadcorrelated} \ee
where  $\sigma_f$ is the overall spatial width of the cloud, $\sigma_{vt}$ is the local velocity spread at any given point $x$ and $T$ is an effective time of evolution. The latter three parameters can be expressed in terms of the parameters $\sigma_{x0}$ and $\sigma_{v0}$ of the initial uncorrelated distribution and the time of evolution as

\be
\sigma_f^2 = \sigma_{x0}^2+\sigma_{v0}^2t^2, \qquad 
\sigma_{vt} = \frac{\sigma_{x0}}{\sigma_f}\sigma_{v0}\,, \qquad
T=t\left(1-\frac{\sigma_{x0}^2}{\sigma_f^2}\right)^{-1}.
\ee

The expression in Eq.\,\ref{eq:quadcorrelated} inside the exponent represents a distribution with correlation between position and velocity, such that at a given point $x$, the width of the velocity distribution is given by $\sigma_{vt}$, which is reduced relative to the initial uncertainty of the velocity by a factor representing the ratio between the initial cloud size and the final cloud size. It is evident that the phase-space volume is conserved during the evolution, as $\sigma_f\sigma_{vt}=\sigma_{x0}\sigma_{v0}$. 

If, at a given time, a distribution has the form of Eq.\,\ref{eq:quadcorrelated}, it is possible to express it as a distribution that started at a time $t$ before this time as an uncorrelated distribution of the form of Eq.\,\ref{eq:uncorrelated}. The variables $\sigma_{x0}$, $\sigma_{v0}$ and $t$ can then be expressed in terms of $\sigma_f$, $\sigma_{vt}$ and $T$. In particular, we find

\be
\sigma_{x0} = \frac{\sigma_{vt} T}{\sqrt{1+\sigma_{vt}^2 T^2/\sigma_f^2}}\,, 
\label{eq:sx0} \ee
and the other two parameters are easily obtained from the latter. 

Let us now consider a situation where an initial uncorrelated distribution has evolved in free space over a time $t_{\rm eff}$ and gave rise to a correlated distribution of the form of an exponent of Eq.\,\ref{eq:quadcorrelated} and then the width of the local velocity distribution $\sigma_{vt}$ has grown due to quick homogeneous heating. This means that after the heating, the parameters $\sigma_f$ and $T$ have remained the same as before the heating, while $\sigma_{vt}$ has grown due to heating. It is now possible to use Eq.\,\ref{eq:sx0} for determining the effective initial uncorrelated distribution that would have led by free propagation to the final distribution after heating. 

If the heating is very strong, such that after the heating $\sigma_{vt}>\sigma_f$, then the projected initial cloud size at the starting point of the evolution is $\sigma_{x0,{\rm eff}}\approx \sigma_f$ and the subsequent evolution of the cloud will be similar to a cloud that starts with the final size of the cloud at the heating time. This means that the heating has erased the position-velocity correlation inside the cloud. Conversely, if the local velocity distribution has not grown considerably during the heating relative to the local velocity uncertainty without heating, then the effect of heating on the consequent evolution is negligible. 

In the case of a squeezed phase-space distribution with squeezing factor $\eta\gg 1$ and a cloud size $\sigma_f$ after the acceleration that is not much larger than the real initial cloud, the local velocity uncertainty is given by $\sigma_{vt}=\eta\sigma_{v0}\cdot (\sigma_{x0}/\eta)/\sigma_f\sim \sigma_{v0}$, which is about the same as the real velocity uncertainty before the squeezing, while the effective initial velocity uncertainty is larger by a factor of $\eta$.  
Let us now assume that after the heating, the local velocity spread grows as $\sigma_{vt}\to \sigma_{vt,{\rm tot}}=\sqrt{\sigma_{vt,0}^2+\sigma_{v,h}^2}$, where $\sigma_{vt,0}$ is the local velocity spread before heating and $\sigma_{v,\rm{heat}}$ is the added spread due to heating. Considering the acceleration procedure and taking the free evolution time before heating to be $t_{\rm eff}$ and assuming $\sigma_f\gg \sigma_{x0,{\rm eff}}$, the effective initial cloud size projected back after heating becomes

\be
\sigma_{x0,{\rm eff}}^h\approx \sigma_{vt,{\rm tot}}t_{\rm eff}=\sqrt{\sigma_{x0,0}^2+\sigma_{v,h}^2t_{\rm eff}^2}=\frac{\sigma_{vt,{\rm tot}}}{\sigma_{vt,0}}\sigma_{x0,0}\,, 
\ee
where $\sigma_{x0,{\rm eff}}^0=\sigma_{x0}/\eta$ is the initial cloud size without heating and $\sigma_{x0,{\rm eff}}^h$ is the projected initial cloud size after heating. Here, we assumed that the ratio between the local velocity spread after heating and before heating is much smaller than the squeezing factor $\eta$. 

\subsection*{Fully quantum derivation of a PSI - wave-packet approach}

Although this work is concerned with thermal atomic clouds with classical evolution, we add here a description of the PSI interferometer from a quantum perspective. The description is based on the wave packet approach of Ref.\cite{japha_unified_2021} and may be extended back to the classical regime by treating a thermal cloud of atoms with a density matrix diagonal in the basis of hermite-Gaussian states.
This approach was not used in the main text of the paper but may provide a starting point for extension to the quantum regime if a BEC is used as the atomic source. 

Consider an initial wave-function $|\psi(t=0)\rangle=\psi({\bf r},t=0)|0\rangle$ in an initial internal state $|0\rangle$ subject to a sequence of Raman pulses $\pi/2 -\pi-\pi/2$  with one arm of the interferometer going through the state sequence $|0\rangle\to|1\rangle\to|0\rangle\to |1\rangle$ and the other through $|0\rangle\to|0\rangle\to |1\rangle\to |1\rangle$. 
In the inertial frame the initial $\pi/2$ pulse applies a momentum transfer $k_{\rm eff}$ in the $\hat{z}$ direction to the first arm. The second pulse ($\pi$ pulse) applies momentum kicks $\mp k_{\rm eff}$ in a rotated direction $\cos\Omega T_R \hat{z}+\sin \Omega T_R\hat{x}$ to the two arms, and the last $\pi/2$ pulse applies the positive momentum to the first arm in the rotated direction $\cos 2\Omega T_R\hat{z}+\sin 2\Omega T_R\hat{x}$. When $\Omega T_R\ll 1$   we can ignore the effect of the rotation in the $\hat{z}$ direction, as $\cos\Omega T_R\approx 1$, and consider only the dynamics along the $\hat{x}$ direction, where we can approximate $\sin\Omega T_R\approx \Omega T_R$. The two classical paths representing the wave-packet centers  end up in the same momentum $p_x=\hbar k_{\rm eff}\Omega T_R$ but in different positions: The first and second arms end up at $x_{\mp}=x_0+2v_0t_R\mp v_R\Omega T_R^2$, where $v_R=\hbar k_{\rm eff}/m$ is the recoil velocity.  The difference between the phases at the wave-packet centers due to the velocity difference in the $x$-direction follows from the  rotation 
\be \delta\phi_{\rm rotation}=\frac{m}{2\hbar}\left[(v_0-v_R\Omega T_R)^2-(v_0+v_R\Omega T_R)^2\right]T_R=-2k_{\rm eff}v_0\Omega T_R^2\,. 
\label{eq:dphi_rotation} \ee
The wave function in the state $|1\rangle$ after the sequence (time $t_f$) is then 
\be \langle 1|\psi(t_f)\rangle=\frac{1}{\sqrt{2}}[\psi_{\sigma}(x-x_-,t_f)+\psi_{\sigma}(x-x_+,t_f)e^{-i\delta\phi}], \ee
where $\delta\phi$ is the phase difference between the two arms, including rotation and gravity contributions, and $\psi_{\sigma}(x,t)$ is the wavefunction (reduced to one dimension for simplicity) after expansion from an initial width $\sigma_0$ to a final width $\sigma(t)$ after a time $t=t_f=T_{\rm ex} $. 

It follows that the final wave function is similar to the result of an expansion starting with two wave-packets centered at a relative distance 
\be d=|x_--x_+|=2v_R\Omega T_R^2\,. \ee
After the expansion this leads to the emergence of a fringe pattern in a manner equivalent to the double-slit experiment. Here the distance $d$ is proportional to the square of the interferometer time and the fringe periodicity in the limit of long expansion time is then given by the far-field limit of the double-slit interference,
\be k_{\Omega}^{PS}=\frac{md}{\hbar T_{\rm ex}}=\frac{2k_{\rm eff}\Omega T_R^2}{T_{\rm ex}}\,, \label{eq:k_PS} \ee
which is identical to the point-source limit of Eq.\,(3) of the main text.

After deriving the point-source limit, we are going to obtain the full details of the signal for a finite-size wave packet expanding for a finite time. 
The evolution of a Gaussian wave-packet in free space, or alternatively a BEC wave-packet in free space or, more generally, in a quadratic potential can be expressed in terms of the evolution of the width $\sigma(t)$. The expression for one dimensional free-space propagation in the frame moving with the center of the wave-packet is given by\cite{castin_bose-einstein_1996,miller_high-contrast_2005}
\be \psi_{\sigma}(x,t)=\frac{\psi_0[x/\lambda(t)]}{\sqrt{\lambda(t)}}\exp\left(\frac{im}{2\hbar}\frac{\dot{\sigma}}{\sigma}x^2\right)\,, 
\label{eq:wp} \ee
where $\psi_0(x)$ is the initial wave-packet wave function and $\lambda(t)=\sigma(t)/\sigma(0)$ is the scaling factor of the wave-packet width. The evolution of the width of a Gaussian wave-packet in free space is given by
\be \sigma(t)=\sqrt{\sigma(0)^2+\frac{\hbar^2 t^2}{4m^2\sigma(0)^2}}=\sqrt{\sigma_{x0}^2+\sigma_{v0}^2t^2}\,, 
\label{eq:sigmat} \ee
with $\sigma_{x0}\equiv \sigma(0)$ and $\sigma_{v0}=\hbar/2m\sigma_{x0}$ for a coherent Gaussian wave-packet, whose momentum uncertainty is proportional to the inverse of the position uncertainty $\sigma(0)$. 

If we take, without loss of generality, $x_0=0$ and $v_0=0$ then the rotational phase $\delta\phi_{\rm rotation}$ of Eq.\,(\ref{eq:dphi_rotation}) vanishes. Here we also ignore the gravitational phase due to the propagation along $z$. The phase difference between the two wave functions centered at $x_{\mp}$ at a given point $x$ follows from the quadratic phases of the expanding Gaussians of the form of Eq.\,(\ref{eq:wp}) and given by
\be \delta\phi(x)=\frac{m\dot{\sigma}}{2\hbar\sigma}[(x-d/2)^2-(x+d/2)^2]=-k_{\Omega}x, \ee
where 
\be k_{\Omega}=\frac{m}{\hbar}\frac{\dot{\sigma}}{\sigma}d=k_{\omega}^{PS}\alpha\,, \ee
where $\alpha=\dot{\sigma}T_{\rm ex}/\sigma$ with $\sigma$ given at $t=T_{\rm ex}$. By taking $\sigma(t)$ to be given in Eq.\,(\ref{eq:sigmat}) we obtain
\be \alpha=\frac{\sigma_{v0}^2T_{\rm ex}^2}{\sigma_{x0}^2+\sigma_{v0}^2T_{\rm ex}^2}=1-\frac{\sigma_{x0}^2}{\sigma_f^2}\,, \ee
which is identical to the definition of $\alpha$ after Eq.\,(4) of the main text.    

The  contrast is  obtained from the amplitude of the interference term
\be \exp\left[-\frac{(x-d/2)^2}{4\sigma_f^2}-\frac{(x+d/2)^2}{4\sigma_f^2}\right]=e^{-x^2/2\sigma_f^2}e^{-d^2/8\sigma_f^2}\,. \ee 
It follows that the contrast is
\be C=e^{-d^2/8\sigma_f^2}=\exp\left[-\frac12 \alpha (k_{\Omega}^{PS}\sigma_{x0})^2\right]\,, 
\label{eq:Contrast} \ee
where we have used $d=(\hbar/m)k_{\Omega}^{PS}T_{\rm ex}=2\sigma_{x0}\sigma_{v0}T_{\rm ex}k_{\Omega}^{PS}$ from Eq.\,(\ref{eq:k_PS}) and the minimum uncertainty relation $m\sigma_{v0}\sigma_{x0}=\hbar/2$. Eq.\,(\ref{eq:Contrast})  is identical with Eq.\,(5) of the main text and we have therefore completed the proof of the identity of the quantum signal with the semi-classical results. 

A thermal cloud could be modeled as a mixture of states. For example, a thermal cloud in an initial harmonic trap is a mixture of hermite-Gaussian states 
\be \rho(x,x')=\sum_n w_n \Phi_n(x)\Phi_n^*(x'), \ee
where  $n$ is the number of nodes in the function and runs from $0$ to $\infty$ and $w_n\propto \exp(-n\hbar\omega/k_BT)$ are the weights (probabilities) of the states $|n\rangle$. The wave function of each of these states is a product of a Gaussian with the same width $\sigma$ for all $n$ and a Hermite polynomial and the evolution of these wave functions in free space follows the same scaling as the state $|n=0\rangle$ analyzed above. 
A thermal Bolzmann distribution yields a Gaussian spatial distribution $\sum_n w_n|\Phi_n(x)|^2\propto\exp(-x^2/2\sigma_T^2)$ with $\sigma_T=\sqrt{k_BT/m\omega^2}$ in a harmonic trap. 
It can be shown that the signal resulting from an interferometric sequence that uses a thermal cloud is then identical to the signal resulting from the coherent wave packet, with the position and momentum widths $\sigma_{x0}$ and $\sigma_{v0}$ replaced by the thermal phase-space widths. These results agree completely with the results that follow from the semi-classical derivation that was used in the main text.

\bibliography{scibib}

\begin{thebibliography}{10}

\bibitem{kasevich_atomic_1991}
M.~Kasevich, S.~Chu, Atomic interferometry using stimulated {Raman}
  transitions. {\it Phys. Rev. Lett.\/} {\bf 67}, 181 (1991).

\bibitem{keith_interferometer_1991}
D.~W. Keith, C.~R. Ekstrom, Q.~A. Turchette, D.~E. Pritchard, An interferometer
  for atoms. {\it Phys. Rev. Lett.\/} {\bf 66}, 2693 (1991).

\bibitem{carnal_youngs_1991}
O.~Carnal, J.~Mlynek, Young{\textquoteright}s double-slit experiment with
  atoms: {A} simple atom interferometer. {\it Phys. Rev. Lett.\/} {\bf 66},
  2689 (1991).

\bibitem{riehle_optical_1991}
F.~Riehle, T.~Kisters, A.~Witte, J.~Helmcke, C.~J. Bord{\'e}, Optical {Ramsey}
  spectroscopy in a rotating frame: {Sagnac} effect in a matter-wave
  interferometer. {\it Phys. Rev. Lett.\/} {\bf 67}, 177 (1991).

\bibitem{cronin_optics_2009}
A.~D. Cronin, J.~Schmiedmayer, D.~E. Pritchard, Optics and interferometry with
  atoms and molecules. {\it Rev. Mod. Phys.\/} {\bf 81}, 1051 (2009).

\bibitem{anderson_sagnac_1994}
R.~Anderson, H.~R. Bilger, G.~E. Stedman,
  {\textquoteleft}{\textquoteleft}{Sagnac}{\textquoteright}{\textquoteright}
  effect: {A} century of {Earth}-rotated interferometers. {\it American Journal
  of Physics\/} {\bf 62}, 975 (1994).

\bibitem{garrido_alzar_compact_2019}
C.~L. Garrido~Alzar, Compact chip-scale guided cold atom gyrometers for
  inertial navigation: {Enabling} technologies and design study. {\it AVS
  Quantum Science\/} {\bf 1}, 014702 (2019).

\bibitem{savoie_interleaved_2018}
D.~Savoie, M.~Altorio, B.~Fang, L.~A. Sidorenkov, R.~Geiger, A.~Landragin,
  Interleaved atom interferometry for high-sensitivity inertial measurements.
  {\it Sci. Adv.\/} {\bf 4}, eaau7948 (2018).

\bibitem{geiger_continuous_2016}
I.~Dutta, D.~Savoie, B.~Fang, B.~Venon, C.~Garrido~Alzar, R.~Geiger,
  A.~Landragin, Continuous {Cold}-{Atom} {Inertial} {Sensor} with 1 nrad / sec
  {Rotation} {Stability}. {\it Phys. Rev. Lett.\/} {\bf 116}, 183003 (2016).

\bibitem{gillot_stability_2014}
P.~Gillot, O.~Francis, A.~Landragin, F.~Pereira Dos~Santos, S.~Merlet,
  Stability comparison of two absolute gravimeters: optical versus atomic
  interferometers. {\it Metrologia\/} {\bf 51}, L15 (2014).

\bibitem{bertoldi_aedge_2021}
A.~Bertoldi, {\it et~al.\/}, {AEDGE}: {Atomic} experiment for dark matter and
  gravity exploration in space. {\it Exp Astron\/} {\bf 51}, 1417 (2021).

\bibitem{bassi_way_2022}
A.~Bassi, L.~Cacciapuoti, S.~Capozziello, S.~Dell{\textquoteright}Agnello,
  E.~Diamanti, D.~Giulini, L.~Iess, P.~Jetzer, S.~K. Joshi, A.~Landragin, C.~L.
  Poncin-Lafitte, E.~Rasel, A.~Roura, C.~Salomon, H.~Ulbricht, A way forward
  for fundamental physics in space. {\it npj Microgravity\/} {\bf 8}, 49
  (2022).

\bibitem{elliott_quantum_2023}
E.~R. Elliott, {\it et~al.\/}, Quantum gas mixtures and dual-species atom
  interferometry in space. {\it Nature\/} {\bf 623}, 502 (2023).

\bibitem{abend_technology_2023}
S.~Abend, {\it et~al.\/}, Technology roadmap for cold-atoms based quantum
  inertial sensor in space. {\it AVS Quantum Science\/} {\bf 5}, 019201 (2023).

\bibitem{gustavson_rotation_2000}
T.~L. Gustavson, A.~Landragin, M.~A. Kasevich, Rotation sensing with a dual
  atom-interferometer {Sagnac} gyroscope. {\it Class. Quantum Grav.\/} {\bf
  17}, 2385 (2000).

\bibitem{durfee_long-term_2006}
D.~S. Durfee, Y.~K. Shaham, M.~A. Kasevich, Long-{Term} {Stability} of an
  {Area}-{Reversible} {Atom}-{Interferometer} {Sagnac} {Gyroscope}. {\it Phys.
  Rev. Lett.\/} {\bf 97}, 240801 (2006).

\bibitem{canuel_six-axis_2006}
B.~Canuel, F.~Leduc, D.~Holleville, A.~Gauguet, J.~Fils, A.~Virdis, A.~Clairon,
  N.~Dimarcq, C.~J. Bord{\'e}, A.~Landragin, P.~Bouyer, Six-{Axis} {Inertial}
  {Sensor} {Using} {Cold}-{Atom} {Interferometry}. {\it Phys. Rev. Lett.\/}
  {\bf 97}, 010402 (2006).

\bibitem{gauguet_characterization_2009}
A.~Gauguet, B.~Canuel, T.~L{\'e}v{\`e}que, W.~Chaibi, A.~Landragin,
  Characterization and limits of a cold-atom {Sagnac} interferometer. {\it
  Phys. Rev. A\/} {\bf 80}, 063604 (2009).

\bibitem{stevenson_sagnac_2015}
R.~Stevenson, M.~R. Hush, T.~Bishop, I.~Lesanovsky, T.~Fernholz, Sagnac
  {Interferometry} with a {Single} {Atomic} {Clock}. {\it Phys. Rev. Lett.\/}
  {\bf 115}, 163001 (2015).

\bibitem{gautier_accurate_2022}
R.~Gautier, M.~Guessoum, L.~A. Sidorenkov, Q.~Bouton, A.~Landragin, R.~Geiger,
  Accurate measurement of the {Sagnac} effect for matter waves. {\it Science
  Advances\/} {\bf 8}, eabn8009 (2022).

\bibitem{janvier_compact_2022}
C.~Janvier, V.~M{\'e}noret, B.~Desruelle, S.~Merlet, A.~Landragin, F.~Pereira
  Dos~Santos, Compact differential gravimeter at the quantum projection-noise
  limit. {\it Phys. Rev. A\/} {\bf 105}, 022801 (2022).

\bibitem{gauguet_off-resonant_2008}
A.~Gauguet, T.~E. Mehlst{\"a}ubler, T.~L{\'e}v{\`e}que, J.~Le~Gou{\"e}t,
  W.~Chaibi, B.~Canuel, A.~Clairon, F.~P. Dos~Santos, A.~Landragin,
  Off-resonant {Raman} transition impact in an atom interferometer. {\it Phys.
  Rev. A\/} {\bf 78}, 043615 (2008).

\bibitem{stockton_absolute_2011}
J.~K. Stockton, K.~Takase, M.~A. Kasevich, Absolute {Geodetic} {Rotation}
  {Measurement} {Using} {Atom} {Interferometry}. {\it Phys. Rev. Lett.\/} {\bf
  107}, 133001 (2011).

\bibitem{berg_composite-light-pulse_2015}
P.~Berg, S.~Abend, G.~Tackmann, C.~Schubert, E.~Giese, W.~P. Schleich, F.~A.
  Narducci, W.~Ertmer, E.~M. Rasel, Composite-{Light}-{Pulse} {Technique} for
  {High}-{Precision} {Atom} {Interferometry}. {\it Phys. Rev. Lett.\/} {\bf
  114}, 063002 (2015).

\bibitem{yao_calibration_2018}
Z.-W. Yao, S.-B. Lu, R.-B. Li, J.~Luo, J.~Wang, M.-S. Zhan, Calibration of
  atomic trajectories in a large-area dual-atom-interferometer gyroscope. {\it
  Phys. Rev. A\/} {\bf 97}, 013620 (2018).

\bibitem{dickerson_multiaxis_2013}
S.~M. Dickerson, J.~M. Hogan, A.~Sugarbaker, D.~M.~S. Johnson, M.~A. Kasevich,
  Multiaxis {Inertial} {Sensing} with {Long}-{Time} {Point} {Source} {Atom}
  {Interferometry}. {\it Phys. Rev. Lett.\/} {\bf 111}, 083001 (2013).

\bibitem{chen_single-source_2019}
Y.-J. Chen, A.~Hansen, G.~W. Hoth, E.~Ivanov, B.~Pelle, J.~Kitching, E.~A.
  Donley, Single-{Source} {Multiaxis} {Cold}-{Atom} {Interferometer} in a
  {Centimeter}-{Scale} {Cell}. {\it Phys. Rev. Applied\/} {\bf 12}, 014019
  (2019).

\bibitem{avinadav_rotation_2020}
C.~Avinadav, D.~Yankelev, M.~Shuker, O.~Firstenberg, N.~Davidson, Rotation
  sensing with improved stability using point-source atom interferometry. {\it
  Phys. Rev. A\/} {\bf 102}, 013326 (2020).

\bibitem{krzyzanowska_matter-wave_2023}
K.~A. Krzyzanowska, J.~Ferreras, C.~Ryu, E.~C. Samson, M.~G. Boshier,
  Matter-wave analog of a fiber-optic gyroscope. {\it Phys. Rev. A\/} {\bf
  108}, 043305 (2023).

\bibitem{jia_dual_2024}
W.~Jia, P.~Yan, S.~Wang, Y.~Feng, A dual atomic interferometric inertial sensor
  utilizing transversely cooled atomic beams. {\it 2024 IEEE International
  Symposium on Inertial Sensors and Systems (INERTIAL)\/} pp. 1--4 (2024). Doi:
  10.1109/INERTIAL60399.2024.10502115.

\bibitem{ammann_delta_1997}
H.~Ammann, N.~Christensen, Delta {Kick} {Cooling}: {A} {New} {Method} for
  {Cooling} {Atoms}. {\it Phys. Rev. Lett.\/} {\bf 78}, 2088 (1997).

\bibitem{kovachy_matter_2015}
T.~Kovachy, J.~M. Hogan, A.~Sugarbaker, S.~M. Dickerson, C.~A. Donnelly,
  C.~Overstreet, M.~A. Kasevich, Matter {Wave} {Lensing} to {Picokelvin}
  {Temperatures}. {\it Phys. Rev. Lett.\/} {\bf 114}, 143004 (2015).

\bibitem{luan_realization_2018}
T.~Luan, Y.~Li, X.~Zhang, X.~Chen, Realization of two-stage crossed beam
  cooling and the comparison with {Delta}-kick cooling in experiment. {\it
  Review of Scientific Instruments\/} {\bf 89}, 123110 (2018).

\bibitem{dupays_delta-kick_2021}
L.~Dupays, D.~C. Spierings, A.~M. Steinberg, A.~Del~Campo, Delta-kick cooling,
  time-optimal control of scale-invariant dynamics, and shortcuts to
  adiabaticity assisted by kicks. {\it Phys. Rev. Research\/} {\bf 3}, 033261
  (2021).

\bibitem{pandey_atomtronic_2021}
S.~Pandey, H.~Mas, G.~Vasilakis, W.~Von~Klitzing, Atomtronic {Matter}-{Wave}
  {Lensing}. {\it Phys. Rev. Lett.\/} {\bf 126}, 170402 (2021).

\bibitem{peters_high-precision_2001}
A.~Peters, K.~Y. Chung, S.~Chu, High-precision gravity measurements using atom
  interferometry. {\it Metrologia\/} {\bf 38}, 25 (2001).

\bibitem{sagnac_1914}
G.~Sagnac, Effet tourbillonnaire optique. la circulation de l'{\'e}ther
  lumineux dans un interf{\'e}rographe tournant. {\it J. Phys. Theor. Appl.\/}
  {\bf 4}, 177 (1914).

\bibitem{mcguirk_large_2000}
J.~M. McGuirk, M.~J. Snadden, M.~A. Kasevich, Large {Area} {Light}-{Pulse}
  {Atom} {Interferometry}. {\it Phys. Rev. Lett.\/} {\bf 85}, 4498 (2000).

\bibitem{chiow_102_2011}
S.-w. Chiow, T.~Kovachy, H.-C. Chien, M.~A. Kasevich, 102 $\hbar$ k {Large}
  {Area} {Atom} {Interferometers}. {\it Phys. Rev. Lett.\/} {\bf 107}, 130403
  (2011).

\bibitem{kovachy_quantum_2015}
T.~Kovachy, P.~Asenbaum, C.~Overstreet, C.~A. Donnelly, S.~M. Dickerson,
  A.~Sugarbaker, J.~M. Hogan, M.~A. Kasevich, Quantum superposition at the
  half-metre scale. {\it Nature\/} {\bf 528}, 530 (2015).

\bibitem{plotkin-swing_three-path_2018}
B.~Plotkin-Swing, D.~Gochnauer, K.~E. McAlpine, E.~S. Cooper, A.~O. Jamison,
  S.~Gupta, Three-{Path} {Atom} {Interferometry} with {Large} {Momentum}
  {Separation}. {\it Phys. Rev. Lett.\/} {\bf 121}, 133201 (2018).

\bibitem{rudolph_large_2020}
J.~Rudolph, T.~Wilkason, M.~Nantel, H.~Swan, C.~M. Holland, Y.~Jiang, B.~E.
  Garber, S.~P. Carman, J.~M. Hogan, Large {Momentum} {Transfer} {Clock} {Atom}
  {Interferometry} on the 689 nm {Intercombination} {Line} of {Strontium}. {\it
  Phys. Rev. Lett.\/} {\bf 124}, 083604 (2020).

\bibitem{parker_measurement_2018}
R.~H. Parker, C.~Yu, W.~Zhong, B.~Estey, H.~M{\"u}ller, Measurement of the
  fine-structure constant as a test of the {Standard} {Model}. {\it Science\/}
  {\bf 360}, 191 (2018).

\bibitem{gebbe_twin-lattice_2021}
M.~Gebbe, J.-N. Siem{\ss}, M.~Gersemann, H.~M{\"u}ntinga, S.~Herrmann,
  C.~L{\"a}mmerzahl, H.~Ahlers, N.~Gaaloul, C.~Schubert, K.~Hammerer, S.~Abend,
  E.~M. Rasel, Twin-lattice atom interferometry. {\it Nat. Commun.\/} {\bf 12},
  2544 (2021).

\bibitem{li_high_2021}
J.~Li, G.~R.~M. Da~Silva, W.~C. Huang, M.~Fouda, J.~Bonacum, T.~Kovachy, S.~M.
  Shahriar, High {Sensitivity} {Multi}-{Axes} {Rotation} {Sensing} {Using}
  {Large} {Momentum} {Transfer} {Point} {Source} {Atom} {Interferometry}. {\it
  Atoms\/} {\bf 9}, 51 (2021).

\bibitem{dubetsky_sequential_2023}
B.~Dubetsky, Sequential large momentum transfer exploiting rectangular {Raman}
  pulses. {\it Phys. Rev. A\/} {\bf 108}, 063308 (2023).

\bibitem{siems_large-momentum-transfer_2023}
J.-N. Siemß, F.~Fitzek, C.~Schubert, E.~M. Rasel, N.~Gaaloul, K.~Hammerer,
  Large-momentum-transfer atom interferometers with $\mu$rad-accuracy using
  {Bragg} diffraction. {\it Phys. Rev. Lett.\/} {\bf 131}, 033602 (2023).

\bibitem{goerz_robust_2023}
M.~H. Goerz, M.~A. Kasevich, V.~S. Malinovsky, Robust {Optimized} {Pulse}
  {Schemes} for {Atomic} {Fountain} {Interferometry}. {\it Atoms\/} {\bf 11},
  36 (2023).

\bibitem{li_spin-squeezing-enhanced_2023}
J.~Li, G.~R.~M. Da~Silva, S.~Kain, J.~Bonacum, D.~D. Smith, T.~Kovachy, S.~M.
  Shahriar, Spin-squeezing-enhanced dual-species atom interferometric
  accelerometer employing large momentum transfer for precision test of the
  equivalence principle. {\it Phys. Rev. D\/} {\bf 108}, 024011 (2023).

\bibitem{louie_robust_2023}
G.~Louie, Z.~Chen, T.~Deshpande, T.~Kovachy, Robust atom optics for {Bragg}
  atom interferometry. {\it New J. Phys.\/} {\bf 25}, 083017 (2023).

\bibitem{shahriar_analytical_2017}
G.~W. Hoth, B.~Pelle, J.~Kitching, E.~A. Donley, {\it Slow Light, Fast Light,
  and Opto-Atomic Precision Metrology X\/}, S.~M. Shahriar, J.~Scheuer, eds.
  (SPIE, 2017), vol. 10119, p. 1011908.

\bibitem{li_sensitivity_2024}
J.~Li, T.~Kovachy, J.~Bonacum, S.~M. Shahriar, Sensitivity of a
  {Point}-{Source}-{Interferometry}-{Based} {Inertial} {Measurement} {Unit}
  {Employing} {Large} {Momentum} {Transfer} and {Launched} {Atoms}. {\it
  Atoms\/} {\bf 12} (2024).

\bibitem{foster_method_2002}
G.~T. Foster, J.~B. Fixler, J.~M. McGuirk, M.~A. Kasevich, Method of phase
  extraction between coupled atom interferometers using ellipse-specific
  fitting. {\it Opt. Lett.\/} {\bf 27}, 951 (2002).

\bibitem{sugarbaker_enhanced_2013}
A.~Sugarbaker, S.~M. Dickerson, J.~M. Hogan, D.~M.~S. Johnson, M.~A. Kasevich,
  Enhanced {Atom} {Interferometer} {Readout} through the {Application} of
  {Phase} {Shear}. {\it Phys. Rev. Lett.\/} {\bf 111}, 113002 (2013).

\bibitem{chen_robust_2020}
Y.-J. Chen, A.~Hansen, M.~Shuker, R.~Boudot, J.~Kitching, E.~A. Donley, Robust
  inertial sensing with point-source atom interferometry for interferograms
  spanning a partial period. {\it Opt. Express\/} {\bf 28}, 34516 (2020).

\bibitem{romero-isart_coherent_2017}
O.~Romero-Isart, Coherent inflation for large quantum superpositions of
  levitated microspheres. {\it New J. Phys.\/} {\bf 19}, 123029 (2017).

\bibitem{yuce_quantum_2021}
C.~Yuce, Quantum inverted harmonic potential. {\it Phys. Scr.\/} {\bf 96},
  105006 (2021).

\bibitem{weiss_large_2021}
T.~Weiss, M.~Roda-Llordes, E.~Torrontegui, M.~Aspelmeyer, O.~Romero-Isart,
  Large {Quantum} {Delocalization} of a {Levitated} {Nanoparticle} {Using}
  {Optimal} {Control}: {Applications} for {Force} {Sensing} and {Entangling}
  via {Weak} {Forces}. {\it Phys. Rev. Lett.\/} {\bf 127}, 023601 (2021).

\bibitem{ullinger_logarithmic_2022}
F.~Ullinger, M.~Zimmermann, W.~P. Schleich, The logarithmic phase singularity
  in the inverted harmonic oscillator. {\it AVS Quantum Science\/} {\bf 4},
  024402 (2022).

\bibitem{neumeier_fast_2024}
L.~Neumeier, M.~A. Ciampini, O.~Romero-Isart, M.~Aspelmeyer, N.~Kiesel, Fast
  quantum interference of a nanoparticle via optical potential control. {\it
  Proc. Natl. Acad. Sci. U.S.A.\/} {\bf 121} (2024).

\bibitem{rozenman_observation_2024}
G.~G. Rozenman, F.~Ullinger, M.~Zimmermann, M.~A. Efremov, L.~Shemer, W.~P.
  Schleich, A.~Arie, Observation of a phase space horizon with surface gravity
  water waves. {\it Commun Phys\/} {\bf 7}, 165 (2024).

\bibitem{grimm_optical_2000}
R.~Grimm, M.~Weidemuller, Y.~B. Ovchinnikov, Optical dipole traps for neutral
  atoms. {\it Advances in Atomic Molecular and Optical Physics\/} {\bf 42}, 95
  (2000).

\bibitem{castin_bose-einstein_1996}
Y.~Castin, R.~Dum, Bose-{Einstein} {condensates} in {time} {dependent} {traps}.
  {\it Phys. Rev. Lett.\/} {\bf 77}, 5315 (1996).

\bibitem{chaudhuri_realization_2006}
S.~Chaudhuri, S.~Roy, C.~S. Unnikrishnan, Realization of an intense cold {Rb}
  atomic beam based on a two-dimensional magneto-optical trap: {Experiments}
  and comparison with simulations. {\it Phys. Rev. A\/} {\bf 74}, 023406
  (2006).

\bibitem{japha_unified_2021}
Y.~Japha, Unified model of matter-wave-packet evolution and application to
  spatial coherence of atom interferometers. {\it Phys. Rev. A\/} {\bf 104},
  053310 (2021).

\bibitem{miller_high-contrast_2005}
D.~E. Miller, J.~R. Anglin, J.~R. Abo-Shaeer, K.~Xu, J.~K. Chin, W.~Ketterle,
  High-contrast interference in a thermal cloud of atoms. {\it Phys. Rev. A\/}
  {\bf 71}, 043615 (2005).

\end{thebibliography}

\bibliographystyle{Science}

\end{document}